\documentclass[fleqn,usenatbib]{mnras}

\usepackage{newtxtext,newtxmath}

\usepackage[T1]{fontenc}
\usepackage{ae,aecompl}

\DeclareRobustCommand{\VAN}[3]{#2}
\let\VANthebibliography\thebibliography
\def\thebibliography{\DeclareRobustCommand{\VAN}[3]{##3}\VANthebibliography}

\usepackage{graphicx}	
\usepackage{amsmath}	
\usepackage{comment}
\usepackage{threeparttable}
\usepackage{longtable}
\usepackage{ragged2e}

\title[Transient cores in high mass star forming regions]{Transient protostellar cores in high mass star forming regions revealed by time-resolved synthetic imaging of dust emission}

\author[Camilo H. Pe\~naloza et al.]{
Camilo H. Pe\~naloza,$^{1}$\href{https://orcid.org/0000-0003-4312-9966}{\includegraphics[scale=0.5]{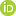}}  
Rowan J. Smith,$^{1}$\href{https://orcid.org/0000-0002-0820-1814}{\includegraphics[scale=0.5]{orcid_16x16.jpeg}}\thanks{E-mail: rjs22@st-andrews.ac.uk} 
Claudia J.\ Cyganowski,$^{1}$\href{https://orcid.org/0000-0001-6725-1734}{\includegraphics[scale=0.5]{orcid_16x16.jpeg}} 
Gwenllian M. Williams,$^{2}$\href{https://orcid.org/0000-0001-5933-2147}{\includegraphics[scale=0.5]{orcid_16x16.jpeg}} 
\newauthor
Michael C.\ Logue,$^{1}$\href{https://orcid.org/0009-0005-7600-7190}{\includegraphics[scale=0.5]{orcid_16x16.jpeg}} 
Todd R.\ Hunter$^{3}$\href{https://orcid.org/0000-0001-6492-0090}{\includegraphics[scale=0.5]{orcid_16x16.jpeg}} 
and Jiancheng Feng$^{4}$\href{https://orcid.org/0009-0000-3311-0159}{\includegraphics[scale=0.5]{orcid_16x16.jpeg}} 
\\
$^{1}$University of St Andrews, SUPA, School of Physics \& Astronomy, North Haugh, St Andrews,UK, KY16 9SS\\
$^{2}$Department of Physics, Aberystwyth University, Ceredigion, Cymru, SY23 3BZ, UK\\
$^{3}$National Radio Astronomy Observatory, 520 Edgemont Rd, Charlottesville, VA, 22903, USA\\
$^{4}$Shanghai Astronomical Observatory, Chinese Academy of Sciences, 80 Nandan Road, Shanghai, 200030, China\\
}

\date{Accepted XXX. Received YYY; in original form ZZZ}

\pubyear{2025}

\begin{document}
\label{firstpage}
\pagerange{\pageref{firstpage}--\pageref{lastpage}}
\maketitle

\begin{abstract}
The connection between dense gas cores and their infant protostars is key to understanding how stars form in molecular clouds. 
In this paper we 
investigate the properties, persistence, and protostellar content of cores that would be identified by a dendrogram analysis of 1.3\,mm ALMA images.
We use a time
series of synthetic images produced by post-processing a simulation of star formation in a massive globally collapsing clump,
with \textsc{polaris} to calculate dust radiative transfer and CASA to generate synthetic ALMA data.  
Identifying sinks in the simulation with protostars, we find that most dendrogram-identified cores do not contain any protostars, with many cores being transient features associated with clumpy flow along feeder filaments. 
Cores with protostars generally host $\leq$3, and protostellar mass is not strongly correlated with the mass of the parent cores due to their transience and shifting boundaries.        
Calculating observationally-relevant intensity-weighted average temperatures for all cores, we find that even at early times the core temperature distribution spans tens of Kelvin, and its width increases with time.  
The 1.3\,mm peak and integrated intensity of the brightest mm core do not increase monotonically as the most massive associated protostar grows, indicating it cannot be assumed that brighter mm sources host more massive protostars.  
Leveraging the time domain, we test observational properties that have been proposed as potential evolutionary indicators and find that only the total 1.3\,mm flux density of the region, the total 1.3\,mm flux density
in cores, and the number of cores show strong, statistically significant correlation with time.
\end{abstract}




\section{Introduction}\label{sec:intro}

The identification and characterisation of structures in interferometric (sub)millimetre continuum images are fundamental to studies of highly clustered high-mass star-forming regions, and have been for nearly two decades \citep[e.g.][]{Hunter06,Cyganowski07,Zhang07,Brogan09,Bontemps10,Palau13,Wang14,Beuther15,Traficante15,Zhang15,Brogan16,Henshaw16,Cyganowski17,Csengeri17,Sanhueza17,Beuther18,Lu20,Liu22,Coletta_almagal3}.  Indeed, the first examples of precursors to Trapezium-like clusters ("proto-Trapezia") were identified in early Submillimeter Array (SMA) and IRAM Plateau de Bure Interferometer (PdBI) images of NGC6334~I, NGC6334~I(N) and W3 IRS5 \citep{Hunter06,Rodon08}.
Many recent studies, including the ALMA-IMF large program \citep{almaimf1,almaimf2,almaimf3,almaimf5,almaimf6,Louvet24}, the ALMA Survey of 70$\mu$m Dark High-mass Clumps in Early Stages (ASHES) \citep{Sanhueza19,Morii23}, the ALMA Evolutionary Study of High Mass Protocluster Formation in the Galaxy (ALMAGAL) \citep{Molinari_almagal1,Coletta_almagal3}, and the ALMA and VLA INvestigations of massive Filaments ANd sTar formation (INFANT) \citep{Cheng24}, focus on using millimetre continuum images to derive the core mass function (CMF) in clustered high-mass star-forming regions as a means of investigating the origin and universality of the stellar initial mass function (IMF) \citep[see also e.g.][]{Cheng18,Liu18,Kong19,Cao21,Takemura21,Suarez21,ONeill21,Takemura23}.  Implicit in these studies -- including comparisons of the mass distributions of pre- and proto-stellar cores \citep[e.g.][]{almaimf5,kong21} -- is an assumption that cores persist as identifiable structures in high-mass, clustered environments over timescales relevant for star formation.

There are reasons to believe that this assumption might not be the case. Unlike the situation in low mass star forming regions, massive star formation is associated with the systemic collapse of the parent molecular cloud \citep[e.g.][]{Peretto07,Motte14,Rigby24}. This collapse funnels gas towards the central objects, allowing them to grow with a high accretion rate and exceed the local Jeans mass within the cloud \citep{Smith09}. \citet{Vazquez19} formalise this process as the `global hierarchical collapse' model under which the molecular cloud clump collapses in an almost pressureless regime, causing filamentary accretion flows to form between the cloud and core scales. Under such a scenario, the gas around the central object will be in a constant state of flux as a result of these accretion streams, meaning that the central gas cores are also likely to be in a state of flux due to ongoing mass accretion onto them \citep{Smith09,Vazquez19,Peretto20,kong21,Rigby21,Rigby24}. The observation that massive star forming regions are associated with Hub Filament systems \citep[e.g.][]{Liu12,Trevino19,Kumar20,Zhou22} is further evidence for the importance of filamentary accretion streams.

Although the dynamical timescales of global hierarchical collapse are short, they are still beyond the lifetime of a human observer.  Therefore, the validity of the assumption that cores persist as identifiable structures can only be tested through numerical simulations.

Several studies have investigated the evolution of dense cores and/or the relationship between the 
CMF and the IMF in "simulation space", i.e. making use of the full 3D information available in numerical simulations \citep[e.g.]
[see also review by \citealt{Offner14}]
{Smith09imf,Gong15,Smullen20,Pelkonen21,Offner22,Offner25}.
In this context, however, "cores" have often been defined based on their potential wells or on instability criteria that are not directly accessible observationally \citep[e.g.][]{Smith09imf,Gong15,Pelkonen21}.\citet{Smullen20} and \citet{Offner22,Offner25} instead identified dense cores in their simulations with dendrograms, which are commonly used to characterise hierarchical structures in observations of star-forming molecular clouds \citep[e.g.][]{Goodman09,Friesen16,Kauffmann17,Henshaw17,Cheng18,Liu18,Rigby18,Williams18,Watkins19,Lu20,ONeill21,Suarez21,Williams22,Williams23,Budaiev23,Peretto23,Morii23}.  
However, since the dendrograms were created from the 3D densities from their simulations \citep{Smullen20,Offner22,Offner25}, the identified structures are not directly comparable to cores identified observationally.

Extracting structures that would be identified as "cores" in observations from simulation snapshots requires synthetic observations that include both radiative transfer and instrumental effects. The importance of synthetic observations for comparing simulations and observations is increasingly well-recognised \citep[e.g. reviews by][]{Haworth18,Rosen20}.  For example, 
synthetic 
high-resolution
(sub)millimetre dust continuum images have been used to study the substructure, stability, and fragmentation of discs around high-mass protostars  
\citep[e.g.][]{Meyer18,Ahmadi19,Jankovic19,Meyer19}.
To date\footnote{While this paper was under review, \citet{Nucara25} published a study of a suite of synthetic observations of massive star-forming clumps at $\sim$7000 AU-resolution, designed to match the ALMA SQUALO survey \citep{Traficante23}.  Considering two or three values for each parameter, \citet{Nucara25} explored the impact of initial clump mass, level of turbulence, and mass-to-magnetic-flux ratio on the degree of fragmentation seen in the synthetic observations, and concluded that magnetic fields most affected fragmentation on the scales studied.}, however, there have been only a few synthetic observational studies of core-scale (sub)millimetre continuum emission in clustered high-mass star-forming regions.  \citet{Smith09} presented synthetic Submillimeter Array (SMA) images of 3 clumps at 3 time-steps from their simulations.  Based on qualitative comparisons, they found that -- at the scales and sensitivities probed by the SMA -- the mm continuum emission becomes dominated by a few (1-2) bright sources at later times \citep{Smith09}.  More recently, \citet{Padoan23} generated synthetic 1.3~mm ALMA images of 12 high-column-density-selected regions in their simulations to quantitatively compare core masses estimated from the synthetic dust maps with those from the 3D simulation data.  While their 12 regions are selected across four simulation snapshots, \citet{Padoan23} focus on mass comparisons rather than the time evolution of the cores.    

\begin{figure*}
    \includegraphics[width=\textwidth,trim={0 0 0 1cm},clip]{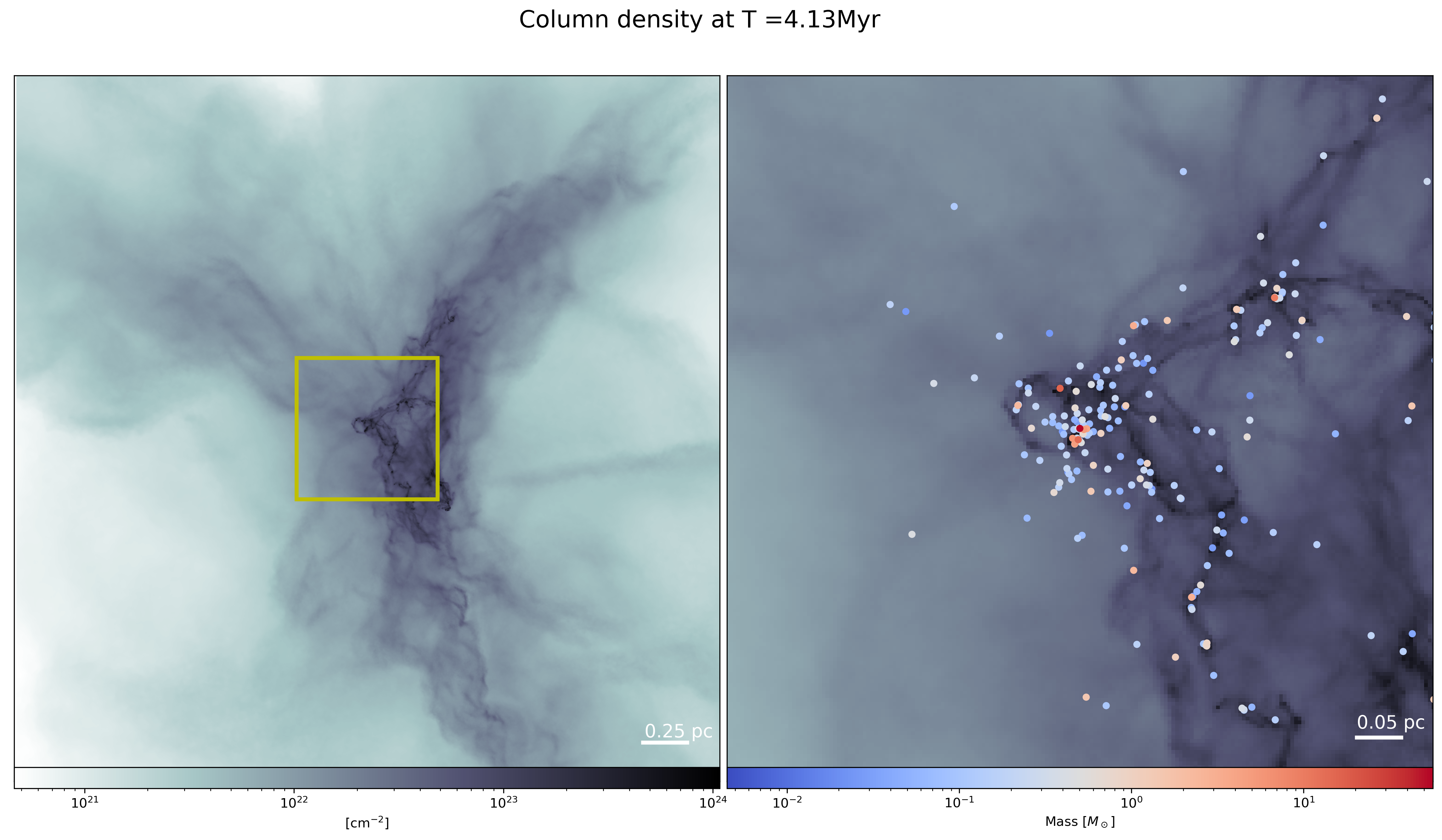}
    \caption{The column density of the cloud at $T=4.135$ Myr, centred on the location of the most massive sink in the snapshot. The left panel shows the large scale filamentary structure of the cloud that creates overdense regions where sink particles form. The right panel shows a zoom view; the field of view shown corresponds to the yellow rectangle in the left panel. 
    In the right panel, positions of sink particles with masses $\geq 0.02 M_{\odot}$ at the end of the simulation are overlaid, colour-coded by their current total mass.}  
    \label{fig:NH}
\end{figure*}

In this paper, we focus on using the time-resolved nature of hydrodynamic simulations to analyse synthetic images including interferometric effects as a function of time.  Using time-resolved synthetic images, we investigate the persistence of observationally-identifiable cores over time, and their relationship to star formation, in high-mass, highly clustered star-forming regions in globally collapsing clouds.  We describe our methods in Section~\ref{sec:methods} and our results in Section~\ref{sec:results}.  We discuss our results in Section~\ref{sec:discuss} and summarise our main conclusions in Section~\ref{sec:conclusions}.

\section{Methods}\label{sec:methods}

\subsection{Hydrodynamical Simulations}\label{sec:sims}

Our simulations use the same initial condition as in \citet{Smith16} but re-simulated such that we reach scales as small as $20$~AU (compared to our previous work where sink particles accreted all gas within a radius of $200$~AU of the sink). By enhancing the resolution beyond that which is typical for simulations of forming star clusters \citep[e.g][]{Lebreuilly25}, we now explicitly model the majority of the protostellar disc surrounding each sink particle. 
(Note that the inner $20$~AU of the disc will still be inside the sink, meaning that the temperature of the inner disc may be underestimated in our simulations.  This should have little effect on the millimetre-wavelength dust emission predicted here, particularly as our post-processing radiative transfer uses a pixel size of 137.5 AU (see Section~\ref{sec:RT}).)
As in \citet{Smith16}, our initial condition is that of a $10^4$ M$_\odot$ cloud of initially atomic gas, whose internal turbulence is initially just sufficient to support the cloud against collapse. As the turbulence decays the cloud collapses (driving new turbulence within the cloud) and a network of filaments and cores forms.

To model the evolution of the gas we use a modified version of the \textsc{Arepo} \citep{Springel10,Pakmor16} code. \textsc{arepo} uses an unstructured Voronoi mesh, which makes it highly versatile and suitable to explore a large dynamical range. This flexibility enables us to model the large scale structure and motions of a molecular cloud while achieving high resolution at the core level. 
Our version of \textsc{Arepo} \citep[e.g.][]{Weinberger20} includes custom modules that capture the physics of the interstellar medium (ISM). To model the heating and cooling of the gas we include time-dependent chemistry to follow the formation of H$_2$ and CO based on the framework of \citet{Nelson97} as described in \citet{Glover12a}. Correctly estimating the attenuation of the Interstellar Radiation Field is important to model the cold gas that will eventually form protostellar cores. We use the \textsc{TreeCol} algorithm described in \citet{Clark12b} to calculate it. Finally, to trace the evolution of protostellar cores we use sink particles (first implemented in \textsc{Arepo} in \citealt{Greif11}) to represent bound, unambiguously collapsing dense gas.

A full description of the simulation setup can be found in \cite{Smith16}; we use the same initial conditions as Cloud 1 in their nomenclature but re-simulated at higher resolution in the dense gas. To ensure that cores and filaments are fully resolved, and to avoid artificial fragmentation, we require that the Jeans length be resolved by 16 resolution elements (note that this criterion is much greater than the minimum requirement of 4 found by \citealt{Truelove97}).

When forming sink particles \cite{Smith16} performed energy checks on gas cells above a number density of $10^7$ cm$^{-3}$ and formed sink particles with an accretion radius of $0.01$ pc when the region was gravitationally bound and had both inward velocities and accelerations. For our new models, we raise the sink creation density to $10^8$ cm$^{-3}$ and decrease the accretion radius to $20$ AU. These adjustments ensure that only the inner accretion disc of a protostellar system would be within the sink and that the sinks represent individual protostellar systems rather than small clusters. There is no feedback tied to the sinks during the hydrodynamical evolution (although radiative transfer is included later, as described in Section \ref{sec:RT}), and consequently the sink particles represent an upper limit to the amount of fragmentation likely within a massive clump. The simulation should be thought of as a numerical experiment aimed at testing how easily low-mass structures, such as protostellar cores, can be recovered within a simulated observation. 

Figure \ref{fig:NH} shows the column density of the cloud at $T=4.135$\,Myr, focusing on the region that contains the most massive sink particle. The region is highly filamentary, with multiple smaller sub-filaments converging on the region where the most massive objects form. Overlaid in the right panel are the positions of the sink particles, with their colours denoting their current total mass. 
As illustrated in Figure~\ref{fig:NH}, many sinks form with a distribution that mostly follows that of the gas, but there are occasions when sinks are ejected from the sub-cluster in which they form and the positions become decoupled.

\subsection{Post-Process Radiative Transfer}\label{sec:rt}
\label{sec:RT}
We use the publicly available radiative transfer code \textsc{polaris} \citep{Reissl16} to model the 1.3~mm dust emission from the cloud for each of 24 time snapshots from the simulation (see Table~\ref{tab:snapshot_stats}). To generate synthetic dust emission maps, \textsc{polaris} calculates the dust heating from user-specified photon-emitting sources, then performs dust radiative transfer, 
iteratively solving for the dust temperature that satisfies radiative equilibrium.  \textsc{polaris} can perform radiative transfer modelling on different grid structures, including the same unstructured Voronoi mesh used in \textsc{arepo}. Using the same grid structure guarantees a one-to-one mapping of the physical quantities and ensures that we do not lose resolution on the smallest scales. 

The yellow rectangle in Figure \ref{fig:NH} shows the extent of the region that is post processed with \textsc{polaris}.  As interferometric observations of high-mass star-forming regions typically target known massive protostars, for each snapshot we centre the post-processed region on the position of the most massive sink particle. For the \textsc{polaris} radiative transfer modelling, we adopt a source distance of $2 {\rm kpc}$ and post-process a $0.8 {\rm pc} \times 0.8 {\rm pc}$ box.  We emphasise that we are not aiming to model any particular high mass star-forming region or to reproduce a specific set of observations, but adopt parameters generally informed by ALMA studies of high mass star-forming regions in the Milky Way (see Section~\ref{sec:intro}).
To ensure the synthetic ALMA images are not limited by the resolution of the \textsc{polaris} output, we generate 1200 pixel$\times$1200 pixel \textsc{Polaris} images (corresponding to a pixel size of 137.5 AU, or 0\farcs069 at 2~kpc).    
 
To calculate the dust heating and radiative transfer, \textsc{polaris} requires as inputs the positions and luminosities of emission sources, in this case (proto)stars. 
As noted in Section~\ref{sec:sims}, our \textsc{arepo} simulations track the position and accretion history of every sink particle at any given time, but do not include radiative feedback. To add this effect in post-processing, we first make use of the pre-main sequence (PMS) data from the MESA Isochrones \& Stellar Tracks (MIST) database \citep{Dotter16,Choi16,Paxton11,Paxton13,Paxton15} to obtain an estimate of the luminosity, $L_{*}$, and protostellar radius, $R_{*}$, of the young stellar object (YSO) that corresponds to the mass of each sink particle, assuming a $50\%$ efficiency of formation as often adopted based on core-to-star measurements at larger scales \citep{Konyves15}. We then calculate the accretion luminosity given this radius by: 

\begin{equation}
    L_{acc}= \frac{G M_{\star} \dot{M}}{R_{\star}}
\end{equation}

\noindent where $M_{\star}$ is the mass of the protostar (here, half the mass of the sink particle), and $\dot{M}$ is the mass accreted between snapshots.

The final luminosity used is then $L = L_{*} + L_{acc}$ which is converted to a temperature given the stellar radius predicted from MIST and used as the input for \textsc{polaris}.
The sink particles are converted into photon emitting sources using the `star' class in \textsc{polaris}. Consequently each is treated as a single point source where the starting direction of the photon packages is random. The wavelength distribution is drawn from $\lambda\in[0.05,2000.0]$~$\mu$m in $100$ wavelength steps using $10^5$ photon packets.

We include as emission sources only sink particles that correspond to an equivalent stellar mass at the end of the simulation of at least $0.01 M_{\odot}$ at $50\%$ efficiency (i.e. M$_{\rm sink,final}\ge0.02 M_{\odot}$).  We note that as this cut is made based on mass at the end of the simulation, there will be sources below this mass threshold in all except the final snapshot. This approach is a compromise between computational efficiency and observational fidelity, since sinks below this threshold are of negligible luminosity. We tested thresholds down to 0.01 solar masses for both early and late snapshots and confirmed the effect of including very low mass sinks was negligible at all times. 
From this point forward, we use "sink mass" to refer to the equivalent stellar mass of each sink, i.e. the value with the 50\% efficiency applied.

For our use case, the other input required for \textsc{polaris} is the dust model.  
For simplicity, we use the dust database files shipped with \textsc{polaris} for spherical grains, which use the optical data from \citet{WD01}, 
for a dust composition of 62.5\% silicate and 37.5\% graphite\footnote{For graphite, we use 25\% \texttt{'graphite\_perpend'} 
and 12.5\% \texttt{'graphite\_parallel'}; see the \textsc{polaris} manual, available at \url{https://github.com/polaris-MCRT/POLARIS}, for further details.}
grains following an MRN power-law size distribution with exponent $-$3.5 \citep[e.g.][]{Fischer94}.
As our aim is to generate synthetic 1.3\,mm images, in order to obtain a dust opacity $\kappa_{\rm 1.3mm}$ similar to the values for dust grains with ice mantles generally adopted for mm cores \citep[e.g.][]{Ossenkopf94,almaimf1}, we set the maximum radius of the grain size distribution to $8\ {\rm \mu m}$, giving $\kappa_{\rm 1.3mm}\approx1\ {\rm cm^2/g}$.  The \textsc{polaris} 1.3\,mm images are shown in Figure~\ref{fig:polaris_images}.

\subsection{Synthetic ALMA Observations}\label{sec:obs}
\label{sec:synthetic_obs}

\begin{figure}
    \centering
    \includegraphics[width=\linewidth]{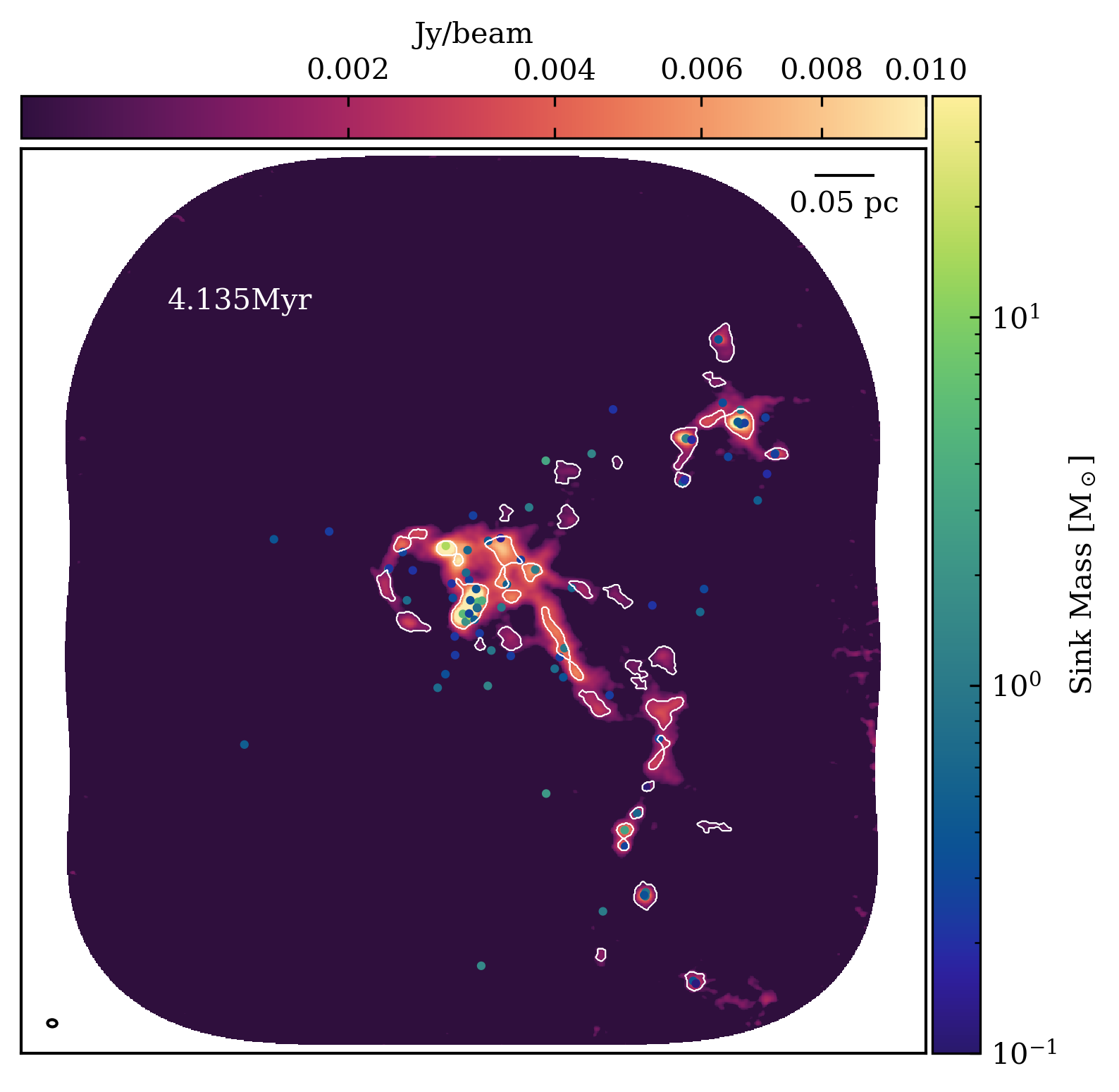}
    \caption{Final synthetic ALMA 1.3\,mm image of the snapshot shown in the right panel of Fig.~\ref{fig:NH}, after post-processing with \textsc{POLARIS} and \textsc{CASA}.  
    The $1.3$ mm dust emission is shown in colourscale, overlaid with markers showing the current masses of sink particles above $1 M_{\odot}$ at the end of the simulation; the contours show cores identified by \textsc{astrodendro} (see Section~\ref{sec:core_ident}).  
    }
    \label{fig:CASA}
\end{figure}

We use the CASA \citep{casa22,McMullin07} \texttt{simobserve} task to generate simulated ALMA measurement sets for each \textsc{polaris} snapshot. 
As we wish to investigate the detectability and persistence of lower-mass cores in the vicinity of an accreting high-mass YSO 
as a function of time, we use antenna configuration 3 of the 12m array and a hexagonal maptype to cover a field size of 1\arcmin\/ with Nyquist pointing spacing. The 30\% response contour of the resulting 27-pointing mosaic has an extent of $\sim$70\arcsec$\times$76\arcsec ($\sim$0.68 pc $\times$0.74 pc at D=2 kpc).    
For our simulated observations, we adopt a total observing time of 45.9 minutes, corresponding to 1.7 minutes per pointing, and an integration time of 6 seconds, which is standard practice for high spectral resolution observations in configuration 3.  

To simulate a synthesised beam and \emph{u,v}-coverage similar to that obtainable for ALMA observations of high mass star-forming regions in the Galactic Plane, we adopt the sky coordinates of G11.92$-$0.61 
\citep[J2000 18$^{\rm h}$13$^{\rm m}$58$\fs$110 $-$18$\degr$54\arcmin22\farcs141, the phase centre of the central mosaic pointing from]
[]{Cyganowski17} 
and centre the observations 1.5 hours after transit.
The effective continuum bandwidth of ALMA observations of high-mass star-forming regions is generally limited by rich line emission from hot cores \citep[e.g.][]{Brogan16}. For consistency, we adopt an effective continuum bandwidth of 1.875 GHz (equivalent to a single ALMA baseband) for all of our simulated observations, though we note that for a given ALMA tuning, the actual effective continuum bandwidth would be expected to depend on source chemistry and evolutionary stage.  
We add thermal noise using the {\tt tsys-atm} option within \texttt{simobserve} assuming 1.796 mm of PWV, the default value from the ALMA Cycle 7 Observing Tool (OT) for the frequency of our simulated observations (231 GHz).

The simulated measurement sets were imaged interactively with the \textsc{CASA} task \texttt{tclean} in CASA 6.2.1.7, using Briggs weighting with a robust parameter of 0.5, multiscale clean, 
and automasking \citep{Kepley20}. The resulting synthetic 1.3~mm continuum images have a synthesised beamsize of 0\farcs81$\times$0\farcs62 ($\sim$1610$\times$1240 AU at D=2 kpc) and the measured rms noise of each snapshot is listed in Table~\ref{tab:snapshot_stats}.  An example synthetic ALMA image is shown in Figure~\ref{fig:CASA}.
Comparison with the raw simulation image of the same snapshot in Figure~\ref{fig:NH} shows that, as expected, much of the diffuse, wispy substructure is missing from the interferometric image but more compact emission from the filaments is recovered (compare also Figure~\ref{fig:24paneldendro} and Figure~\ref{fig:polaris_images}).

\begin{table}
	\centering
	\caption{Snapshot statistics}
    \label{tab:snapshot_stats}
	\begin{threeparttable}
    \label{tab:RMS}
    \begin{tabular}{c|c|c|c|c|c}
        \hline\hline
Snapshot & Time   & $M_{\mathrm{max}}^{a}$ & $\sigma_{\mathrm{rms}}^{b}$ & \# of Cores \\
ID       & (Myr) & (M$_{\odot}$)                              & (mJy\,beam$^{-1}$)                             &             \\ \hline
170      & 3.700  & 0.050                                      & 0.0956                                         & 1           \\
171      & 3.722  & 0.355                                      & 0.0962                                         & 1           \\
172      & 3.743  & 1.010                                      & 0.0970                                         & 2           \\
173      & 3.765  & 1.695                                      & 0.1001                                         & 6           \\
174      & 3.787  & 3.360                                      & 0.1065                                         & 9           \\
175      & 3.809  & 4.625                                      & 0.1095                                         & 9           \\
176      & 3.830  & 5.530                                      & 0.1155                                         & 7           \\
177      & 3.852  & 6.92                                       & 0.1075                                         & 10          \\
178      & 3.874  & 8.045                                      & 0.1295                                         & 11          \\
179      & 3.896  & 9.290                                      & 0.1370                                         & 15          \\
180      & 3.917  & 10.156                                     & 0.1350                                         & 16          \\
181      & 3.939  & 10.980                                     & 0.1325                                         & 24          \\
182      & 3.961  & 11.560                                     & 0.1515                                         & 26          \\
183      & 3.983  & 12.350                                     & 0.1445                                         & 26          \\
184      & 4.004  & 14.825                                     & 0.1555                                         & 18          \\
185      & 4.026  & 17.650                                     & 0.1415                                         & 21          \\
186      & 4.048  & 20.475                                     & 0.1215                                         & 30          \\
187      & 4.070  & 23.905                                     & 0.1265                                         & 23          \\
188      & 4.091  & 26.150                                     & 0.1370                                         & 32          \\
189      & 4.113  & 27.450                                     & 0.1323                                         & 38          \\
190      & 4.135  & 28.105                                     & 0.1340                                         & 43          \\
191      & 4.157  & 28.990                                     & 0.1530                                         & 42          \\
192      & 4.178  & 29.910                                     & 0.1420                                         & 44          \\
193      & 4.200  & 32.365                                     & 0.1175                                         & 52          \\
        \hline\hline
    \end{tabular}
    \begin{tablenotes}
	     \item[$a$] Equivalent stellar mass of the most massive sink in the snapshot, see Section~\ref{sec:RT}.  
	     \item[$b$] Representative rms noise of synthetic ALMA image, measured from emission-free regions within the 50\% mosaic response. Quoted values are for the images used for core identification, i.e. those uncorrected for the primary beam response (see Section~\ref{sec:core_ident}).  
	\end{tablenotes}
	\end{threeparttable}
\end{table}

\subsection{Core identification and properties}\label{sec:core_ident}

We use the \textsc{astrodendro} \textsc{python} package \citep{Rosolowsky08dendro,Robitaille19} to extract structures from our synthetic ALMA images.
The dendrogram algorithm characterises hierarchical structures by stepping through isocontours, and is commonly used to identify structures in 
observational studies of star-forming molecular clouds (see Section~\ref{sec:intro}).
The identification of structures is controlled by
three parameters: the minimum intensity value above which structures are identified ($I_{\mathrm{min}}$), the minimum isocontour separation required to define an independent structure ($\Delta I_{\mathrm{min}}$), and the minimum area of an independent structure (in pixels, $n_{\mathrm{pix}}$). 
Our choice of parameters is motivated by observational studies \citep[e.g.][]{Cheng18,Lu20,Suarez21,Williams22,Williams23,Morii23,Cheng24}, and we adopt
$I_{\mathrm{min}}= 5\sigma_{\mathrm{rms}}$ (where $\sigma_{\mathrm{rms}}$ is the rms noise; see Table~\ref{tab:RMS}), $\Delta I_{\mathrm{min}}=1\sigma_{\mathrm{rms}}$, and $n_{\mathrm{pix}}$=area of the synthesised beam ($\sim$88 pixels). 
While the area of the synthesised beam is the same for all snapshots (as it is determined by the adopted configuration, sky coordinates, and mosaic parameters), the rms of the synthetic images varies from snapshot to snapshot, with those with stronger emission generally exhibiting higher noise.
The rms value used in the dendrogram analysis for each snapshot is included in Table~\ref{tab:RMS}.

Following observational studies, we consider the dendrogram "leaves" (the substructures at the highest level of the isocontour hierarchy, with no further substructure) to be cores and identify structures in images that have not been corrected for the primary beam response \citep[see e.g.][]{Lu20}.
We adopt this approach to counter the effect of the non-uniform response of the primary beam, which results in elevated noise around the edges of the mosaics (see Figure~\ref{fig:24panelsigma}). 
Extracting structures from the uncorrected images prevents strong noise features near the edges of the mosaics from being spuriously identified as sources and minimises the effect of the elevated noise on the masks of structures extracted in the outer regions of the mosaics.
Note that while all leaves extracted by the dendrogram analysis will be referred to as "cores," in reality 
some extracted features may represent regions of filamentary emission; the nature of the extracted "cores" is discussed further in Section~\ref{sec:transient_cores}.

\begin{figure*}
    \centering
    \includegraphics[width=\linewidth]{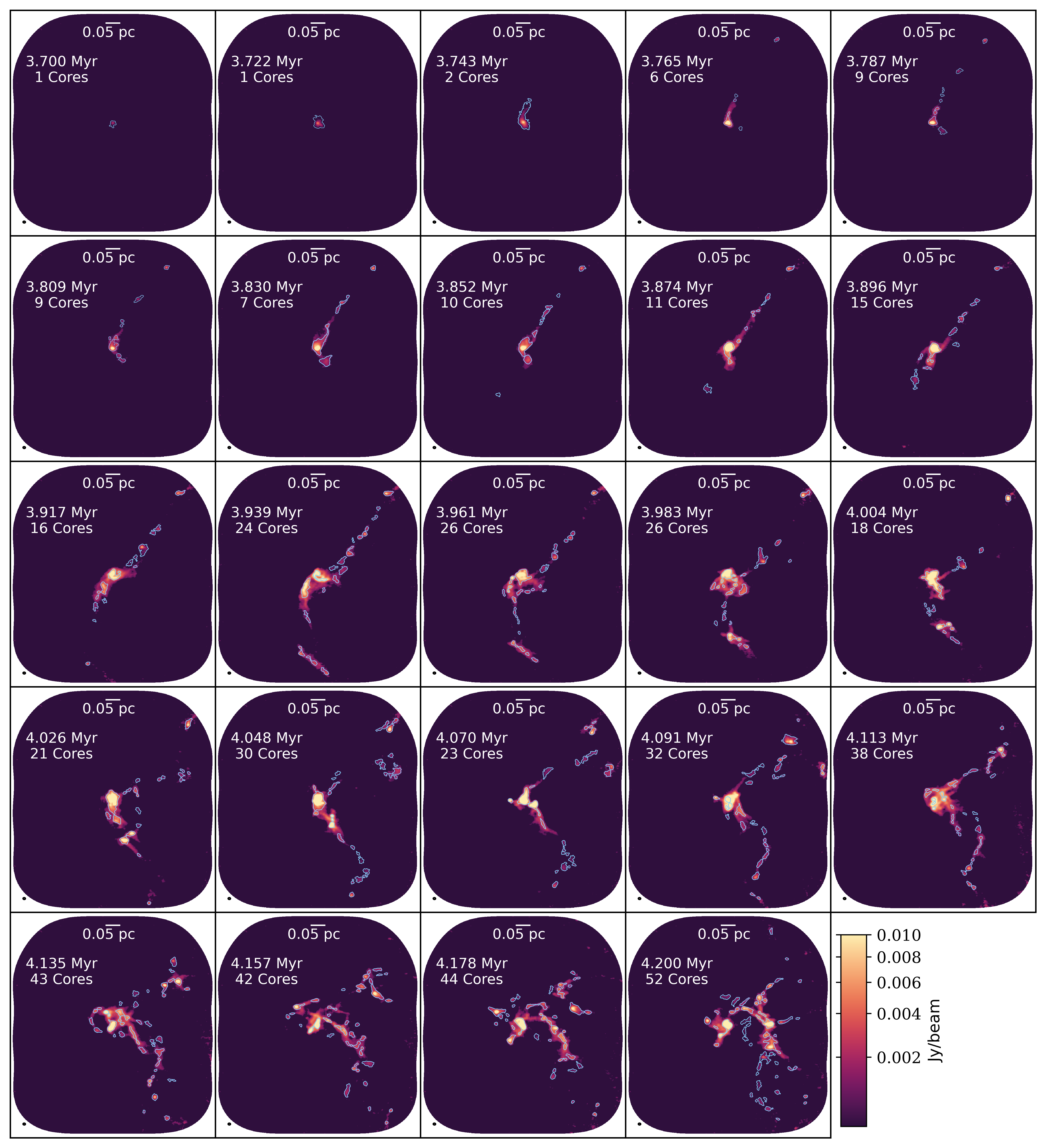}
    \caption{Synthetic ALMA 1.3\,mm continuum images for each snapshot, centred on the position of the most massive sink and with the core boundaries identified with \textsc{astrodendro} (Section~\ref{sec:core_ident}, Table~\ref{tab:properties}) overlaid in blue. The images shown have been corrected for the primary beam response; the colourscale and field of view are the same for all images and the scalebars assume D=2\,kpc (Section~\ref{sec:RT}). In each panel, the synthesised beam is shown at lower left and the time of the snapshot in Myr and number of identified cores are given at upper left.} 
    \label{fig:24paneldendro}
\end{figure*}

Peak intensities and integrated flux densities for the identified cores are then measured within the extracted "leaf" masks using the primary beam corrected images.  We consider all cores within the 30\% level of the mosaic response in our analysis; Figure \ref{fig:24paneldendro} shows the synthetic ALMA image for each snapshot, masked at the 30\% response level and overlaid with contours of the cores extracted with \textsc{astrodendro}.     
Measured properties of the core(s) identified in each snapshot, including    
peak intensity, integrated flux density, and size, are tabulated in
Table \ref{tab:properties}.

\begin{table*}
    \centering
    \fontsize{7.9}{8.2}\selectfont
	\caption{Properties of cores extracted from synthetic ALMA 1.3\,mm images.}
	\label{tab:properties}
    \setlength\tabcolsep{3.69pt}
    \begin{tabular}{ccccccccccccc}
    \hline\hline
    Time$^{a}$             & Core$^{b}$        & Peak intensity$^{c}$          & Integrated flux$^{c}$     & $T_{b}$\,$^{d}$ & $T_{\mathrm{dust}}$\,$^{e}$        & Mass ($T=20$\,K) & Mass ($T=T_{\mathrm{dust}}$) & Mass (sim)$^{f}$            & Source size$^{g}$           & Source size$^{g}$  \\
    (Myr)          & ID          & (mJy\,beam$^{-1}$)            & density (mJy)                     & (K) & (K)                    & (M$_{\odot}$) & (M$_{\odot}$) & (M$_{\odot}$)   & ($\arcsec \times \arcsec$ [P.A. ($^\circ)]$)   & ({\sc au}\,$\times$\,{\sc au}) \\ \hline\hline
    3.700           & 170-1       & 2.18                          & 5.27                      & 1.97     & 11.4               & 0.41  & 0.91  & 2.11           & $2.30 \times 1.90$ [64]                 & $4608 \times 3798$     \\
    3.722           & 171-1       & 4.97                          & 25.00                     & 2.04     & 14.6               & 1.96  & 3.01   & 7.64           & $4.41 \times  3.40$ [108]               & $8811 \times 6800$   \\
    3.743           & 172-1       & 9.14                          & 62.46                     & 2.11     & 18.0               & 4.89  & 5.63  & 15.97           & $10.0 \times  3.66$ [85]                & $20,000 \times 7312$  \\
    3.743           & 172-2       & 0.70                          & 0.75                      & 1.83    & 9.2                & 0.06   & 0.19  & 0.36          & $1.35 \times  0.51$  [151]              & $2690 \times 1026$     \\
    3.765           & 173-1       & 28.24                         & 81.41                     & 3.13    & 28.7                & 6.37  & 4.05  & 7.15           & $2.79 \times  2.32$  [118]              & $5582 \times 4646$     \\
    3.765           & 173-2       & 2.91                          & 3.02                      & 2.48    & 14.8                & 0.24  & 0.36  & 0.75           & $1.63 \times  - $\,\,\,\,\,  [68]       & $3252 \times -$             \\
    3.765           & 173-3       & 2.80                          & 4.66                      & 2.17    & 13.3                & 0.37  & 0.64  & 1.19           & $1.99 \times  0.96$  [$-$146]           & $3987 \times 1917$       \\
    3.765           & 173-4       & 1.24                          & 1.21                      & 2.05    & 10.8                & 0.10  & 0.23  & 0.40           & $1.00 \times  0.58$  [$-$140]           & $1991 \times 1161$       \\
    3.765           & 173-5       & 0.73                          & 0.83                      & 1.85    & 9.9                & 0.06   & 0.18  & 0.37          & $1.07 \times  0.83$  [119]              & $2134 \times 1655$     \\
    3.765           & 173-6       & 0.71                          & 1.05                      & 1.85    & 11.2                & 0.08  & 0.19  & 0.42           & $1.29 \times  0.94$  [94]               & $2580 \times 1874$     \\
    \hline\hline
    \end{tabular}
    \footnotesize
	\begin{flushleft}
	    \vspace{-1ex}
        \justifying
         $^a$ Snapshot time. \\     
	     $^b$ Core ID comprised of the snapshot ID (see Table~\ref{tab:snapshot_stats}) followed by an index number, with cores indexed in each snapshot in order of decreasing peak intensity.  Note that core indices are assigned separately for each snapshot and so \emph{cannot} be taken to reference the same structure across different snapshots (see also Section~\ref{sec:core_numbers}). \\    
	     $^c$ Evaluated within the extracted dendrogram leaf, from primary beam corrected images; see Section~\ref{sec:core_ident}. \\
	     $^d$ Mean Planck brightness temperature within the dendrogram leaf (Section~\ref{sec:core_ident}), calculated from brightness temperature maps generated using the \texttt{tt.brightnessImage} task of \texttt{toddTools}: \href{https://safe.nrao.edu/wiki/bin/view/Main/ToddTools}{https://safe.nrao.edu/wiki/bin/view/Main/ToddTools} \\
	     $^e$ Intensity-weighted average temperature within the dendrogram leaf boundary, calculated from \textsc{Polaris} outputs as described in Section~\ref{sec:core_ident}. \\
      $^f$ Mass within the dendrogram core boundary derived from the \textsc{Polaris} input column density grid prior to radiative transfer, as described in Section~\ref{sec:obs_v_sims}.\\
         $^g$ Calculated from the intensity-weighted second moment from \textsc{astrodendro}, converted to FWHM size and deconvolved from the beam.  Sizes in {\sc au} assume D=2\,kpc (Section~\ref{sec:RT}).  Major or minor axis sizes denoted "$-$" were unable to be deconvolved. The reported position angle, from \textsc{astrodendro}, is measured 
         in degrees North of West; add 90$^{\circ}$ for position angle in degrees East of North.\\ (Only the first ten rows of this table are shown. The full table is available in a machine-readable form in the online journal.)
    \end{flushleft}
\end{table*}

We calculate gas masses for the identified cores from their integrated flux densities assuming optically thin, isothermal dust emission using:
\begin{equation}
	\mathrm{M_{\mathrm{gas}}} =  \frac{d^{2}\,\mathrm{R}\,S_{\nu}}{\kappa_{\nu}\,B_{\nu}(T_{\mathrm{dust}})} \,\, ,
	\label{eq:mdust}
\end{equation} 
\noindent where $d$ is the distance (assumed to be 2\,kpc, Section~\ref{sec:RT}), $R$ is the gas-to-dust mass ratio (assumed to be 100), $S_{\nu}$ is the integrated flux density from Table~\ref{tab:properties}, $\kappa_{\nu}$ is the dust opacity at 1.3\,mm (assumed to be 1\,cm$^2$\,g$^{-1}$, Section~\ref{sec:RT}), and $B_{\nu}(T_{\mathrm{dust}})$ is the Planck function at the dust temperature $T_{\mathrm{dust}}$. 
Observationally constraining the dust temperatures of individual cores is a major challenge \citep[see e.g.][]{Motte25}, hence most observational studies to date adopt a single dust temperature for all cores in a given massive star-forming region, with the assumed temperature sometimes tied to single-dish measurements of the clump-scale dust temperature \citep[e.g.][]{Sanhueza19,Zhang21} or indicators of the region's evolutionary state \citep[e.g.][]{Traficante23, Coletta_almagal3}.   
To explore the implications of this common assumption for studies of core mass functions, we calculate core masses from our synthetic ALMA observations for two cases: 1) a fiducial dust temperature $T_{\mathrm{dust}}$=20\,K for all cores \citep[commonly adopted in observational studies, e.g.][and references therein]{Cyganowski17,Lu20,almaimf1}; and 2) temperatures individually estimated for each core using the \textsc{polaris} outputs.

As noted in Section~\ref{sec:rt}, \textsc{polaris} calculates dust temperatures on a 3D grid as an intermediate step in the radiative transfer.
To generate 2D temperature maps that can be directly compared to our extracted cores, we estimate the observationally-relevant effective dust temperature along the line of sight from the \textsc{polaris} 1.3\,mm intensity and dust opacity maps by assuming local thermodynamic equilibrium and inverting the Planck function.
This approach, like equation~\ref{eq:mdust}, is appropriate for optically thin dust emission. To check this assumption, we calculate a core-average dust optical depth 
\begin{math}
\tau_{\mathrm{dust}} = -\ln\left(1-\frac{T_{b}}{T_{\mathrm{dust}}}\right)   
\end{math}
where, following \citet{Williams23}, $T_{b}$ is the mean brightness temperature of each core measured from the synthetic ALMA 1.3\,mm continuum image (see Table~\ref{tab:properties}).
Across all cores and snapshots, the maximum value of $\tau_{\mathrm{dust}}$ is $\sim$0.27 for case 1 ($T_{\mathrm{dust}}$=20\,K) and $\sim$0.22 for case 2 (individual core dust temperatures), confirming that optically thin dust emission is a reasonable assumption on core scales. We note that, as in observational studies \citep[e.g.][]{Williams23}, this likely underestimates the peak $\tau_{\rm dust}$ of smaller-scale structures.
The temperature assigned to each core is the intensity-weighted average temperature within the \textsc{astrodendro} core boundary, with the intensity-weighting done using the primary-beam-corrected synthetic ALMA 1.3\,mm image.
The intensity-weighted average core temperatures are presented in Table \ref{tab:properties}, along with both core mass estimates, and are discussed in Section~\ref{sec:core_evo}.

\section{Results} \label{sec:results}

\subsection{Synthetic continuum emission}\label{sec:core_numbers}

\begin{figure*}
     \centering
     \includegraphics[width=\linewidth]{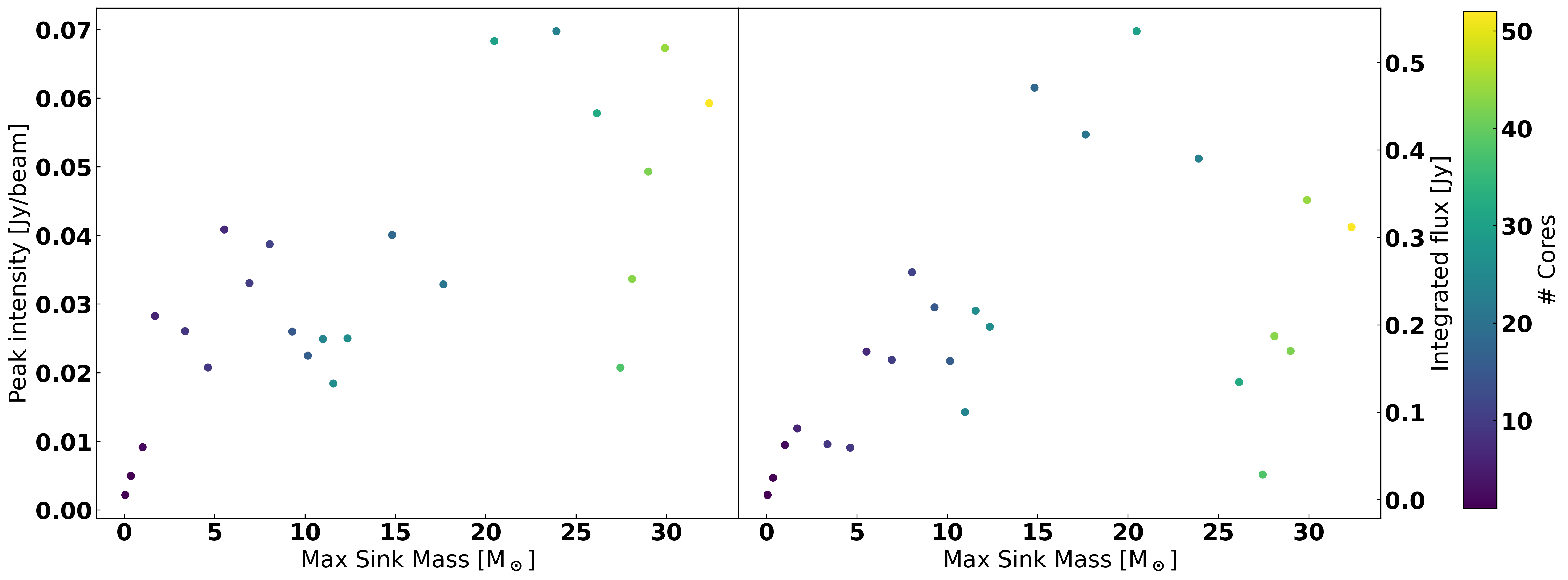}
     \caption{Synthetic ALMA 1.3\, mm peak intensity (left panel) and integrated flux density (right panel) of the brightest mm core in each timestep (the core with the highest peak intensity), plotted against the equivalent stellar mass of the most massive sink in this core. The colour of each point indicates the total number of cores identified in the synthetic image of that snapshot (Table~\ref{tab:snapshot_stats}, see also Figure~\ref{fig:24paneldendro}).  Once the equivalent stellar mass of the most massive sink reaches a few M$_{\odot}$, it has little correlation with the mm peak intensity.}
     \label{fig:PeakInt}
 \end{figure*}

Figure~\ref{fig:24paneldendro} shows the 
synthetic 1.3\,mm ALMA continuum images for all snapshots in time order.  
During the 0.5\,Myr covered by our simulations, the morphology and intensity of the synthetic 1.3\,mm continuum emission evolve significantly. 
Notably, while the intensity of the millimetre continuum emission associated with the central, most massive sink increases markedly over the first few timesteps, this trend does not continue throughout the simulation. Figure~\ref{fig:PeakInt} shows the peak intensity and integrated flux density of the brightest mm core in each timestep (i.e., the core with the highest peak intensity) plotted against the equivalent stellar mass of the most massive sink within this core (assuming 50\% efficiency, see Section~\ref{sec:RT}).
A notable feature of Figure~\ref{fig:PeakInt} is that neither the mm peak intensity nor the mm integrated flux density increase monotonically with sink mass, indicating that it cannot be assumed that brighter mm sources host more massive protostars.
The lack of correlation is most pronounced at higher sink masses, which correspond to later times in the simulation.

Interestingly, while the observable properties of the brightest mm core do not track the mass growth of its most massive sink, Figure~\ref{fig:24paneldendro} does show a striking increase in the overall extent and complexity of the mm continuum emission within the field of view with time.
In the initial snapshots, there is a single core, associated with the sink that will become the most massive sink in the simulation.
As the region evolves, the total 1.3\,mm flux density within the field of view generally increases (see top right panel of Figure \ref{fig:parameters_v_time}) , and more of the filamentary structures present in the cloud (see Section~\ref{sec:sims}) are visible in mm continuum emission.  
This is demonstrated in Figure~\ref{fig:24panelsigma}, which shows the synthetic ALMA images in both colourscale and contours to highlight the extent of detectable emission, as more of the field of view is filled with bright emission over time. 
This can be understood in terms of the density evolution of the region as it undergoes global collapse, combined with the resulting sink formation.
In general, the strength of mm continuum emission depends on both the column density and temperature of the emitting dust: due to global collapse the density of the whole region increases (increasing the dust column density) and new sinks form (and heat their surroundings and so increase the dust temperature), which combine to increase the region's mm continuum emission.

\begin{figure}
    \centering
    \includegraphics[width=\linewidth]{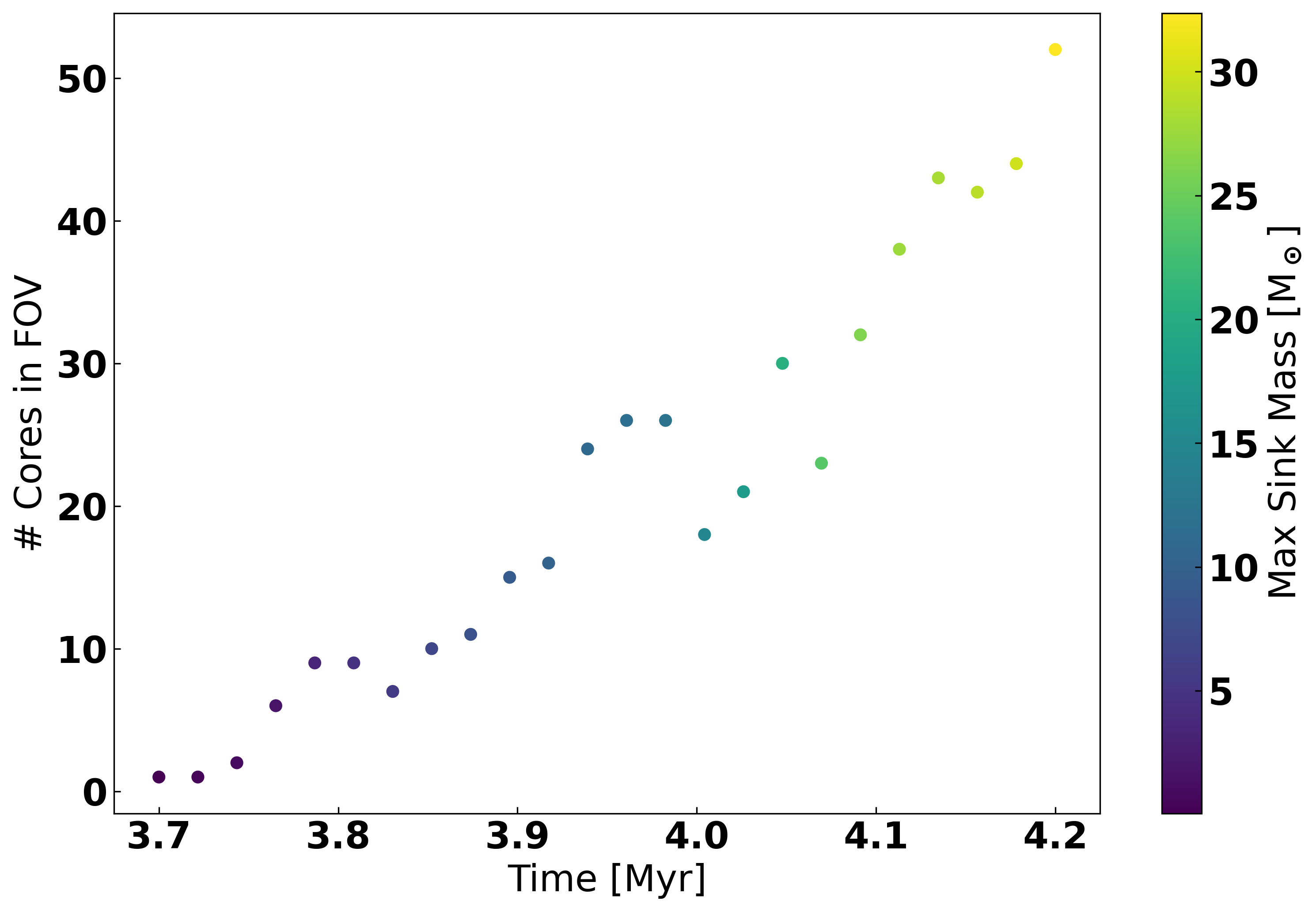}
    \caption{Number of cores identified with {\sc astrodendro} within the 30\% mosaic response (see Section~\ref{sec:core_ident}) plotted against snapshot time. Points are colour coded by the equivalent stellar mass of the most massive sink within the brightest mm core.  Note that this generally corresponds to the most massive sink in the simulation, with the notable exception of T=4.135 Myr, when the most massive sink in the simulation is not within the brightest mm core.}
    \label{fig:CvT}
\end{figure}

\begin{figure}
    \centering
    \includegraphics[width=\linewidth]{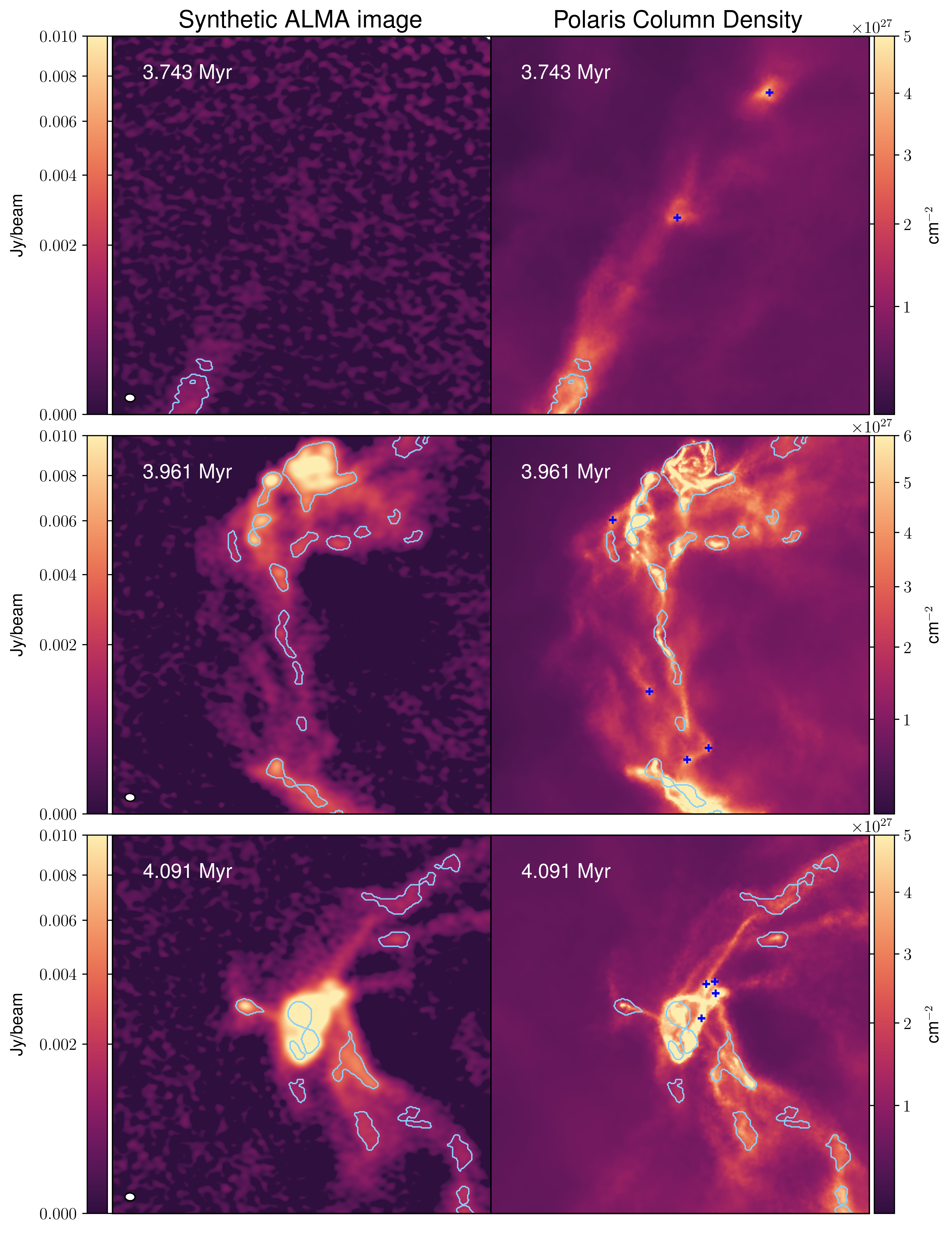}
    \caption{Synthetic 1.3\,mm ALMA images (left) and \textsc{Polaris} column density maps (right) for representative examples of high-column density features in the simulation that do not correspond to extracted mm cores.  All panels are labelled with the snapshot time and show core boundaries identified by \textsc{astrodendro} overlaid in blue contours.  The synthesised beam is shown at lower-left of the ALMA images and selected compact features that do not correspond to extracted cores are highlighted with blue $+$ on the column density maps. }
    \label{fig:IvCD}
\end{figure}

As the region's millimetre continuum emission increases, more cores are identified by the dendrogram analysis. 
Figure \ref{fig:CvT} shows the number of cores identified with {\sc astrodendro} in the field of view as a function of time. Although there are some short temporal variations, there is a clear overall upward trend: the number of cores identified increases with time. Correspondingly, this implies that the number of cores increases with the mass of the central, most massive protostellar object (Table~\ref{tab:RMS}).  Notably, as shown in Figure~\ref{fig:IvCD}, comparison of the synthetic 1.3\,mm ALMA images with column density maps from \textsc{Polaris} prior to radiative transfer shows that the presence of a compact, high column density feature in the simulation is \emph{not} necessarily sufficient for a core to be identified in the corresponding synthetic ALMA image. The identification of cores in the synthetic ALMA images is affected by a complex combination of factors, including the synthesised beamsize (which influences the core boundaries), spatial filtering by the interferometer, and the noise added to the synthetic observations, as well as the underlying dust radiative transfer.

\begin{figure}
    \centering
    \includegraphics[width=\linewidth]{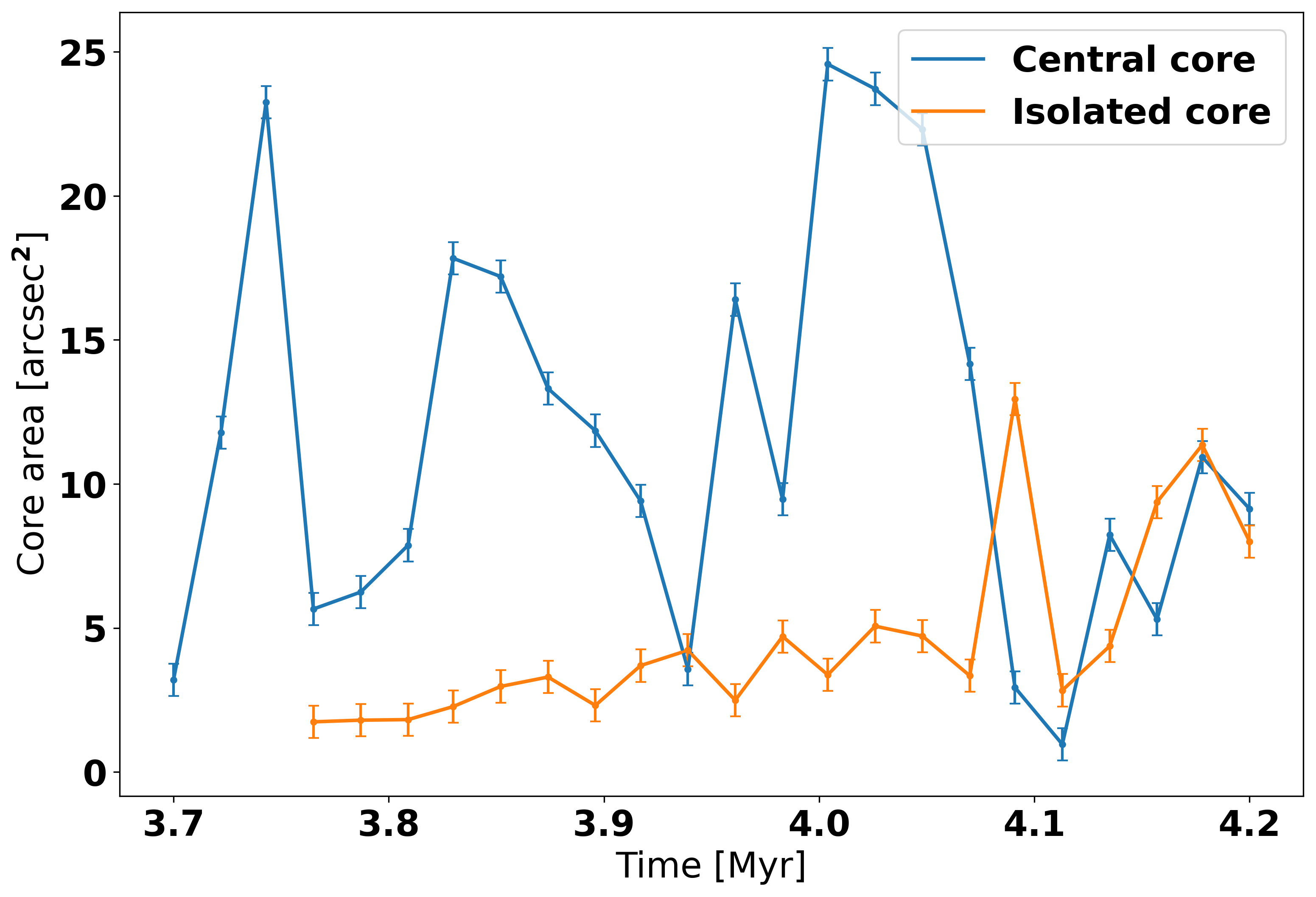}
    \caption{The area of the dendrogram leaf for the two most persistent cores identified in the synthetic images. The error bars are equivalent to the area of the beam.}
    \label{fig:core_area}
\end{figure}

An upward trend in core number with time could reflect ongoing additions to a stable core population: e.g., new cores are identified in each timestep and add to an existing core population identified in previous timesteps.  
The dips seen in Figure~\ref{fig:CvT}, however, indicate that this is not the case.
Close examination of Figure~\ref{fig:24paneldendro} reveals a more complex picture, in which 
the boundaries of the cores, and which parts of the filamentary structure are identified as cores, constantly shift as the region evolves.  
There are very few cores that persist unaltered for multiple snapshots. The most persistent features are the central core associated with the most massive sink and the isolated compact core that appears near the northwest corner of the field of view at T=3.765 Myr.
Figure \ref{fig:core_area} illustrates how the area of these cores change with time. The area contained within the dendrogram boundary of the central core fluctuates wildly with time in response to the rapid timescale over which the emission in the central region changes (see Figure \ref{fig:24panelsigma}). For example, the rapid decrease in core area between T=4.070 Myr and T=4.091 Myr corresponds to the splitting of the central bright emission into multiple cores. In contrast to this, the isolated core's area varies smoothly with little variation between snapshots during the initial phase. This continues until T=4.091 Myr when it joins the top-right filament and merges with another core. After this point the core is no longer isolated and fluctuates in area like the other cores.
As we discuss in Section~\ref{sec:transient_cores}, this complexity 
is linked to the dynamical evolution of the region,
and points to differences in the nature of cores identified in globally collapsing high-mass star-forming regions as compared to isolated low-mass cores. 

\subsection{Core Temperature and Mass Evolution}\label{sec:core_evo}

As noted in Section~\ref{sec:core_ident}, observational studies of core populations generally assume a single dust temperature for all cores identified within a given massive star-forming clump, 
historically commonly 20~K \citep[e.g.][]{Cyganowski17, Lu20,almaimf1}.  Recent studies of large samples have aimed to improve on this assumption by linking the adopted temperature for each region to its clump-scale luminosity-to-mass ratio \citep[L/M, an observationally-accessible indicator of evolutionary state, e.g.][]{Traficante23,Coletta_almagal3}; however, this approach still does not account for variations in the evolutionary stage -- and so temperature -- of individual cores within a given clump. The recent catalogue of mass-averaged temperatures for ALMA-IMF cores by \citet{Motte25} represents a significant advance, but is limited by the resolution of the dust temperature maps, and so uses parametric relations to extrapolate to core-scale temperatures.
Our radiative transfer and synthetic observations allow us to directly probe the range of core temperatures within a single clump and how this distribution evolves as a function of time, and to compare core mass functions (CMFs) derived using these individual core temperatures to CMFs derived assuming a common temperature for all cores.  

\begin{figure*}
        \centering
    \includegraphics[width=\linewidth]{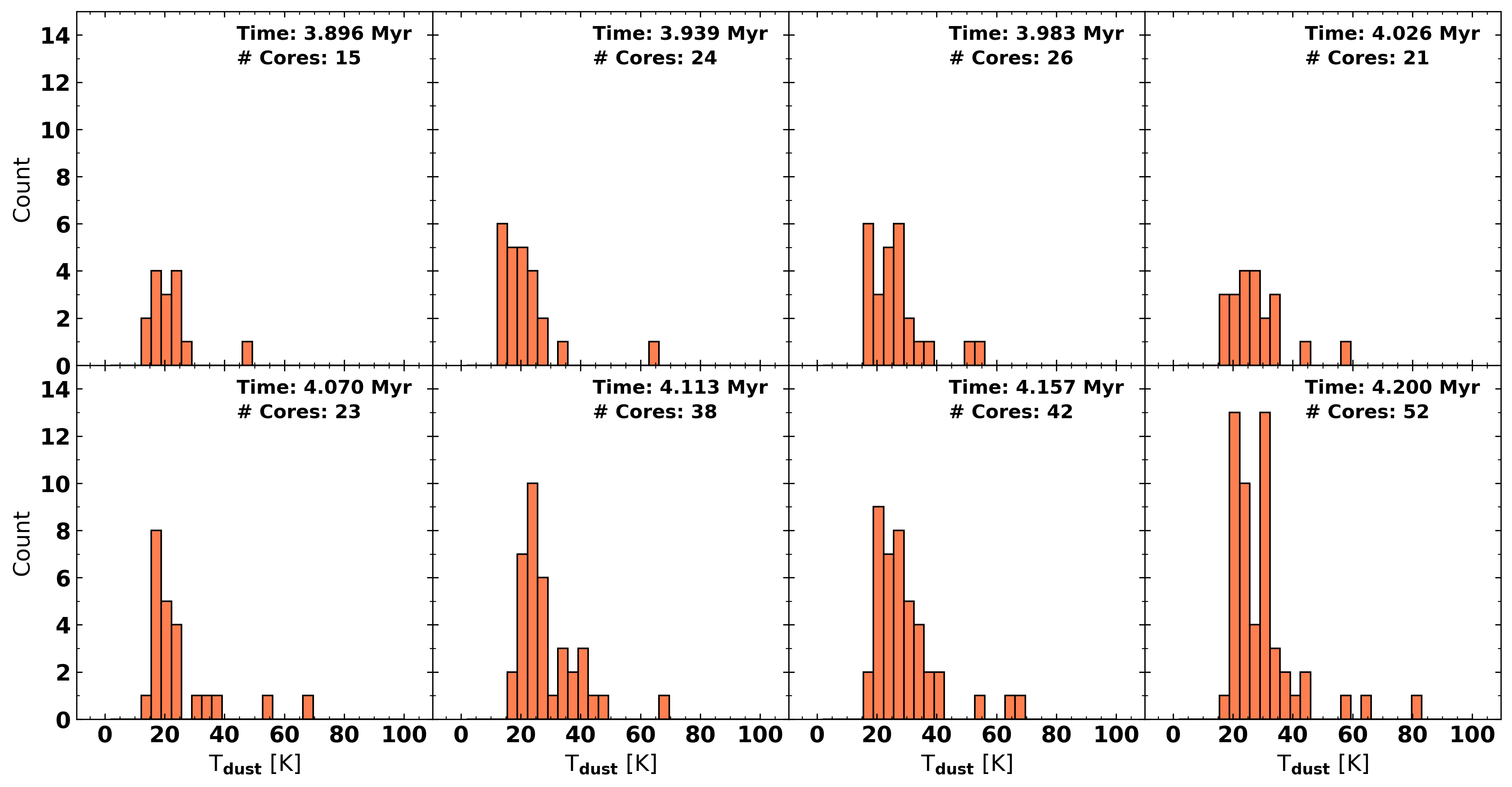}
    \caption{Histograms of the intensity-weighted average core temperature (see Section~\ref{sec:core_ident}) for all cores within the field of view for eight timesteps, ordered from early to late.  Properties of the timesteps are given in Table \ref{tab:snapshot_stats}.}
    \label{fig:Td_hist}
\end{figure*}

Figure~\ref{fig:Td_hist} shows the distribution of intensity-weighted average core temperatures (see Section~\ref{sec:core_ident}) for eight selected timesteps.
As expected, Figure~\ref{fig:Td_hist} shows that the distribution of core temperatures shifts to higher values over time.
Two other notable features also stand out in Figure~\ref{fig:Td_hist}: even at relatively early times (3.896 Myr), the individual core temperatures span a range of 10s of K, and the width of the temperature distribution 
tends to
increase with time. 
Examination of the data for the earliest snapshots (not shown in Figure~\ref{fig:Td_hist} due to the low number of cores) reveals that significant departures from a single-temperature assumption arise as soon as multiple cores are identified.  In the first snapshot with more than one identified core (t=3.743 Myr), when the effective mass of the most massive sink is only $\sim$1M$_{\odot}$, the intensity-weighted temperatures of the two cores differ by $\sim$8.8~K. The first core with an intensity-weighted temperature T$>$20~K appears in the next snapshot, at t=3.765 Myr.

\begin{figure*}
    \centering
    \includegraphics[width=\linewidth]{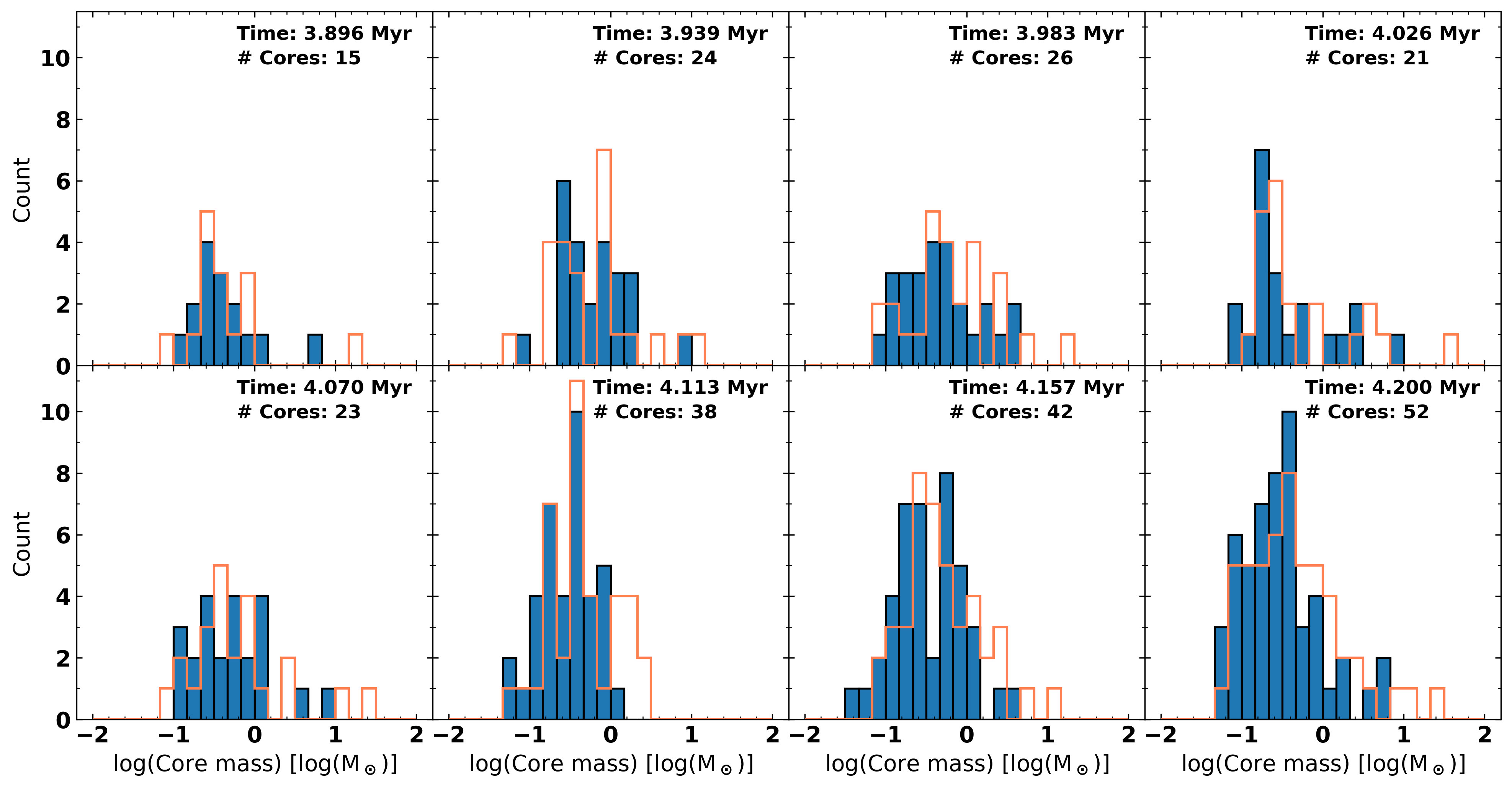}
    \caption{Distribution of derived millimetre core masses within the field of view for the same timesteps shown in Figure~\ref{fig:Td_hist}.  The unshaded orange histograms show masses calculated from the synthetic ALMA 1.3\,mm integrated flux densities using a constant temperature assumption of T$_{\rm dust}$=20~K (case 1 in Section~\ref{sec:core_ident}); the blue histograms show masses calculated using the intensity-weighted individual core temperatures shown in Figure~\ref{fig:Td_hist} (case 2 in Section~\ref{sec:core_ident}).}
    \label{fig:8panelhist}
\end{figure*}

The effect of assuming a constant temperature for all cores on measured CMFs is illustrated in Figure~\ref{fig:8panelhist}, which presents a comparison of CMFs for the two cases described in Section~\ref{sec:core_ident}: (1) a fiducial dust temperature T$_{\rm dust}$=20~K for all cores and (2) individual intensity-weighted core temperatures.
Figure~\ref{fig:8panelhist} clearly demonstrates that, as expected, the constant-temperature assumption overestimates the masses of the highest-mass cores, impacting the shape of the upper end of the CMF.  This happens because the most massive cores are significantly warmer than the adopted temperature of T$_{\rm dust}$=20~K (compare Figure~\ref{fig:Td_hist}). 
We note that while assuming a different (higher) dust temperature
for later snapshots 
\citep[analogous to e.g.][]{Traficante23,Coletta_almagal3} 
would reduce the values in their constant-temperature histograms, it would do so across the board, meaning that the masses of the most massive cores would still be overestimated relative to the rest of the core population.
Notably, a constant temperature assumption also impacts the low mass end and overall shape of the CMF, as cores with T$<$20~K have their masses underestimated in our constant-temperature case.  This effect is most pronounced at earlier times: the fraction of identified cores with T$>$20~K generally increases with time, with some fluctuations, and exceeds 90\% in the final five snapshots (from t=4.113~Myr).

Considering the individual-core-temperature histograms (in blue), Figure~\ref{fig:8panelhist} shows that while
there are a few massive cores, the majority of the core population is low mass, and the
core mass distribution does not evolve monotonically from low to high mass cores over time. 
We refrain from fitting a slope to the high mass end of the core mass function \citep[e.g.][]{Motte18a} because our field of view was selected to contain the most massive sink in the simulated molecular cloud, and the high-mass slope of the CMF would likely become steeper if a wider area was considered.

\begin{figure*}
    \centering
    \includegraphics[width=\linewidth]{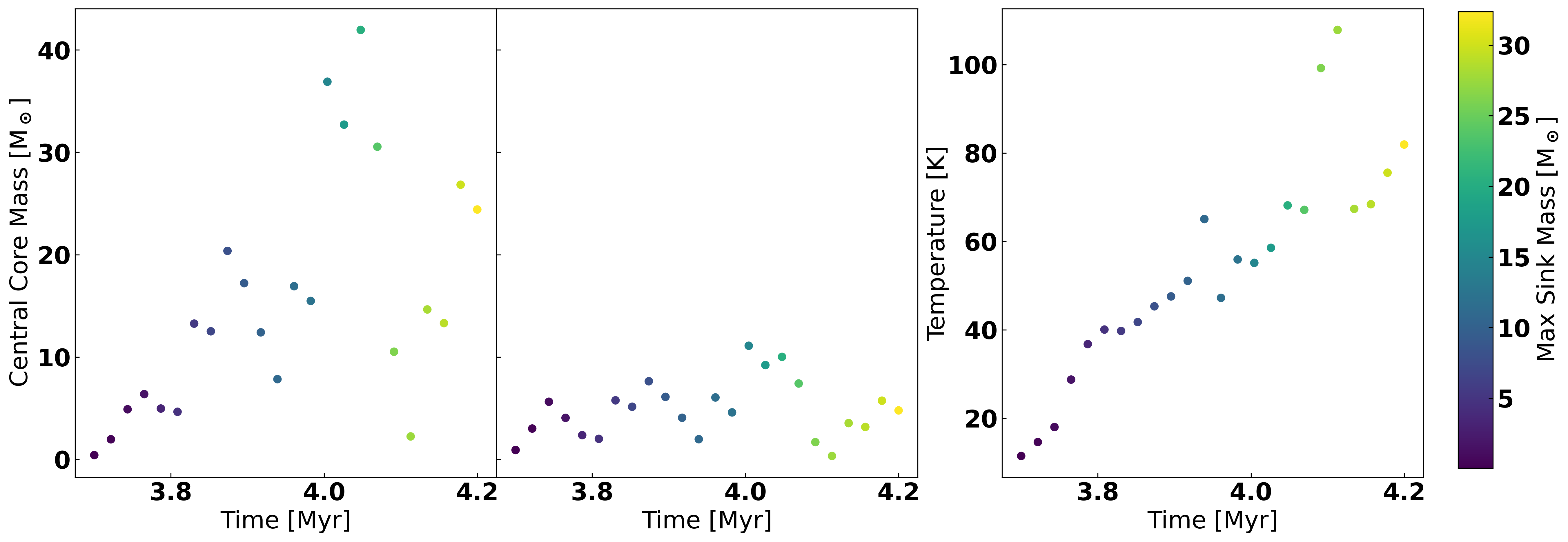}
    \caption{Mass of the brightest mm core in each snapshot calculated using (left:) T$_{\rm dust}$=20~K and (middle:) the core's intensity-weighted average temperature, which is shown at (right).  All panels are plotted as a function of snapshot time.  Points are colour coded by the equivalent stellar mass of the most massive sink within this core. This generally corresponds to the most massive sink in the simulation, with the notable exception of T=4.135\,Myr, when the most massive sink is not within the brightest mm core.}
    \label{fig:CvM}
\end{figure*}

For the brightest mm core in each snapshot (the core shown in Figure~\ref{fig:PeakInt}), Figure \ref{fig:CvM} shows the millimetre-derived mass as a function of time for both temperature cases, alongside the time evolution of the core's intensity-weighted average temperature.
As illustrated by Figure~\ref{fig:CvM}, the adopted temperature dramatically affects the conclusion that would be reached about the mass evolution of this central core.
In the T$_{\rm dust}$=20~K case, the core mass scales with the measured integrated flux density, and so appears to generally increase until t=4.048\,Myr (albeit not monotonically, see also above), reaching a mass of $>$40M$_{\odot}$.  In contrast, when masses are derived using the intensity-weighted individual core temperature, 
the core's mass fluctuates but the overall trend is largely flat with time, and the core's mass peaks at $\sim$11M$_{\odot}$. 
The reason for these differences can be clearly seen in the right-hand panel of  Figure~\ref{fig:CvM}, which shows that the intensity-weighted average core temperature increases strongly -- though again not monotonically -- with time.

\subsection{Cores and Sinks} \label{sec:core2sink}
The level of core fragmentation at various scales, and its potential correlation with core mass or evolutionary state, have been a focus of numerous recent observational studies \citep[e.g.][]{Ishihara24,Yoo25,Yang25,Kao25,Coletta25}.
Having identified cores in our synthetic ALMA images in a manner similar to observational studies, we can use the information on sink locations from the original simulation to investigate the protostellar content of the mm-identified cores. Each sink in the simulation has a unique ID number and so it is trivial to track them over time. We do not track the cores in the same way, since, as will be discussed in Section~\ref{sec:transient_cores}, the cores are not in general long-lived objects. Instead we compare sink masses to the mass of the core that those sink(s) are associated with in a single snapshot. To do this we simply assign a sink to a core when the core's boundary encloses the sink's position on the projected emission map.

\begin{figure}
            \centering
    \includegraphics[width=\linewidth]{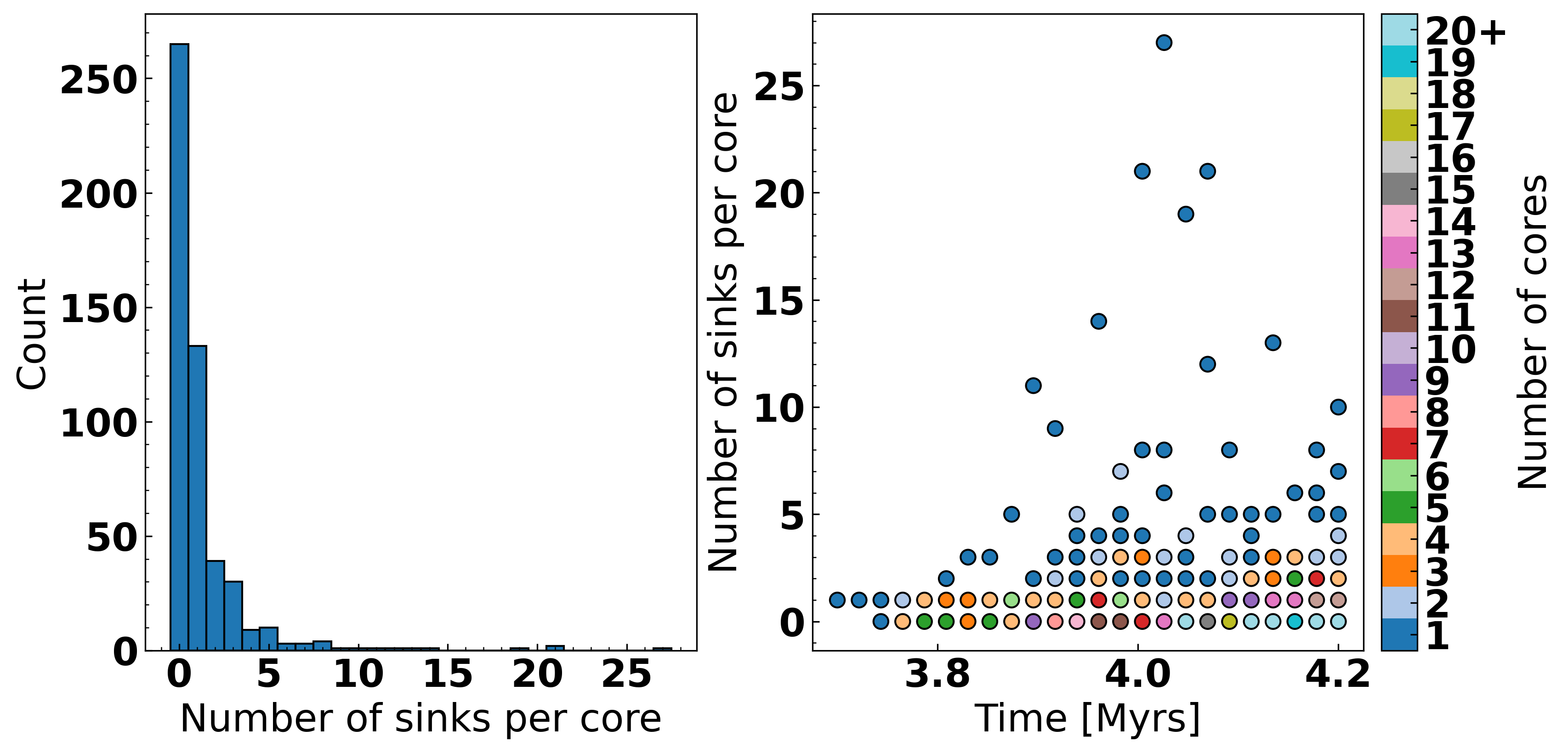}
    \caption{\textit{Left} The combined distribution of number of sinks per core in the simulation over all analysed timesteps. \textit{Right} The distribution of the number of sinks per core at each timestep. The colours of the points indicate how many cores contain a given number of sinks at each time. The vast majority of mm cores contain three or fewer sinks within their boundaries.}
    \label{fig:sink_number}
\end{figure}

The left panel of Figure~\ref{fig:sink_number} shows a histogram of the distribution of the number of sinks per core, summed over all analysed timesteps.  The most striking feature of this plot is the strong peak in the distribution at zero: most of the cores identified in the synthetic mm images with {\sc astrodendro} do not contain a sink within their boundaries. 
This is consistent with different parts of the filamentary structure being identified as cores as the region evolves (Section~\ref{sec:core_numbers}), and with many of these "cores" being transient features that do not host star formation.
A time-resolved view is shown in the right panel of Figure~\ref{fig:sink_number}, which plots how many cores contain a given number of sinks 
for each timestep.  Of the cores that host sinks, most are associated with a small number of sinks ($n_{\rm{sink}}\le 3$), as might be expected for a core forming a single or small-n multiple star system.
In our simulation, it is rare for mm cores to be associated with large numbers of sinks: there is never more than one core enclosing $>$10 sinks in any timestep (Figure~\ref{fig:sink_number}) and this core with many sinks is the central, brightest mm core.

\begin{figure}
    \centering
    \includegraphics[width=\linewidth]{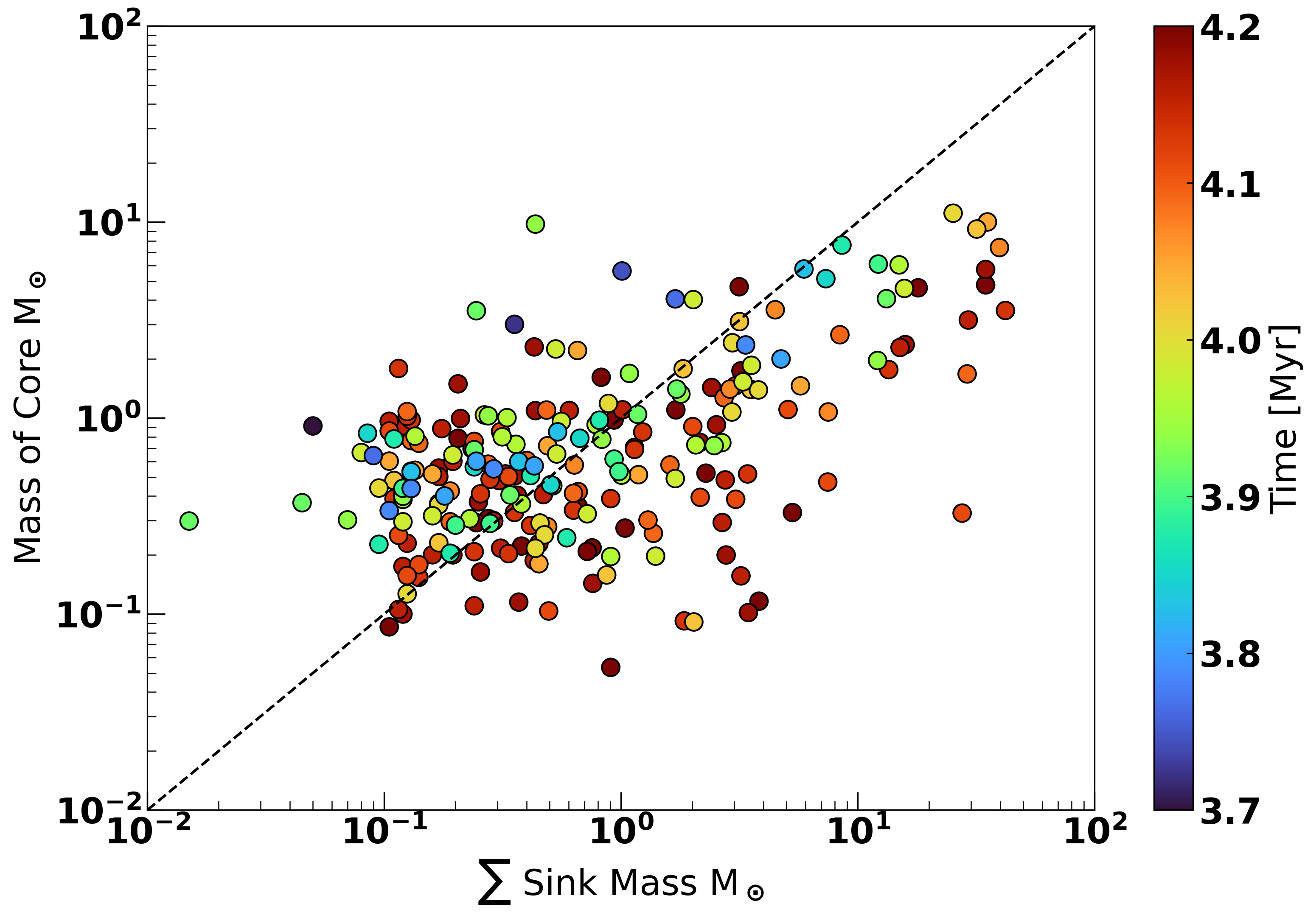}
    \caption{The total equivalent stellar mass in sinks within the boundaries of each mm core, plotted against the mass of the parent core calculated using its intensity-weighted average temperature. The dashed line shows a 1-1 correspondence and the colour of the points shows the time at which the masses were measured. The core masses are higher than the total equivalent stellar mass of the associated sinks only for low total equivalent stellar masses.} 
    \label{fig:CvSM}
\end{figure}

Figure \ref{fig:CvSM} shows the sum of the equivalent stellar mass of all the sinks that fall within the boundary of each core, plotted against the core's mass 
at each timestep in the simulation. As discussed in Section~\ref{sec:core_evo}, a constant-temperature assumption overpredicts the masses of high-mass cores and underpredicts the masses of some low-mass cores; we therefore use our case 2 core masses, calculated using the individual intensity-weighted average core temperatures, for Figure \ref{fig:CvSM} and subsequent analysis.

There is some correlation between the observationally derived mass of a core and that of its stellar population (Pearson correlation coefficient=0.66 with P-value=3.97e-31).
However, Figure \ref{fig:CvSM} shows that the scatter in the trend is at least an order of magnitude. 
The line shows a one-to-one correspondence between the observationally derived core mass and the total equivalent stellar mass of the associated sinks. 
At lower total equivalent stellar masses many points lie above this line, indicating more mass in the core than the sinks, as would be expected for cores at an earlier evolutionary stage.
At higher total equivalent stellar masses, however, the picture is different: above a total equivalent stellar mass of $\sim$2\,M$_{\odot}$, the vast majority of points lie on or below the 1:1 line. 
In the literature a constant efficiency factor is often used to convert between the instantaneous core mass function and the final stellar mass \citep[e.g.][]{Alves07}. Figure \ref{fig:CvSM} shows that this is a poor description even in a statistical sense for cores in globally collapsing massive star forming clumps.

\begin{figure}
    \centering
    \includegraphics[width=\linewidth]{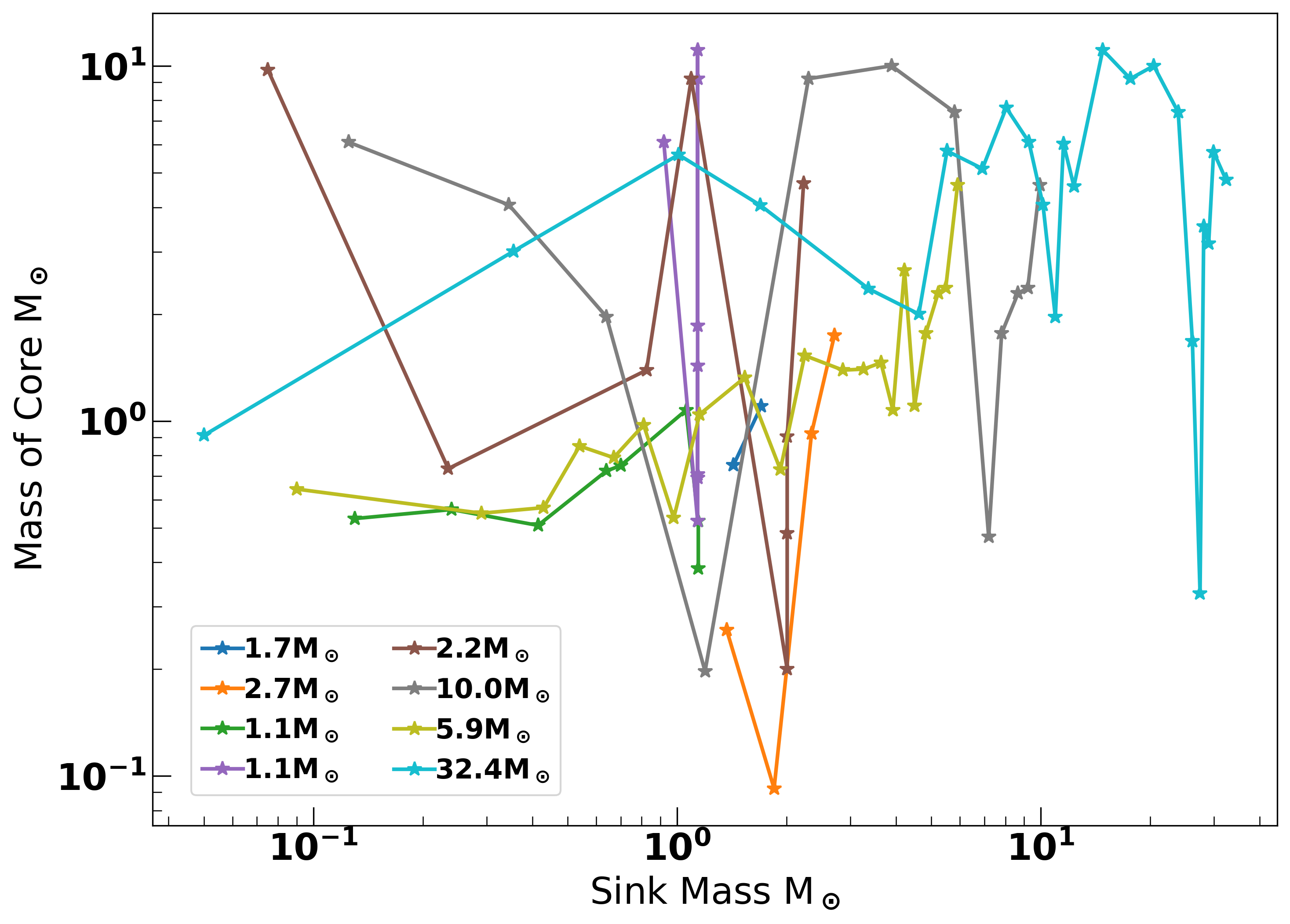}
    \caption{Evolution of the natal core mass of selected sinks. We select sinks that are above $1 M_{\odot}$ at the end of the simulation, identify the core each sink is in at each timestep, and calculate the core's mass using its intensity-weighted average temperature. Each line represents a different sink and is labelled by the equivalent stellar mass of the sink at the end of the simulation.}
    \label{fig:maxSvC}
\end{figure}

In Figure \ref{fig:maxSvC} we explore the natal core environments of different sinks. As the mm cores cannot in general be tracked from timestep to timestep (Section~\ref{sec:core_numbers}), we select an illustrative sample of sinks with an equivalent stellar mass $>$1\,M$_{\odot}$ at the end of the simulation and track the mass of the core in which each sink is embedded throughout its evolution. Most of the sinks exhibit a random walk through this parameter space. 
The only sink to show a steady correlation between sink and core mass is that with a final equivalent stellar mass of 5.9M$_{\odot}$. 
 An investigation of the path of this sink shows it is associated with the core at the top right of the panels in Figure \ref{fig:24paneldendro}, that is relatively isolated during its collapse to the centre. An inspection of the boundaries of this core in Figure~\ref{fig:24paneldendro}, and the core area in Figure~\ref{fig:core_area}, reveal that these remain relatively steady throughout the evolution and lack the extreme fluctuations seen in more crowded environments.

\section{Discussion}\label{sec:discuss}

\subsection{Core Mass Fidelity}\label{sec:obs_v_sims}
\begin{figure}
    \centering
    \includegraphics[width=\linewidth]{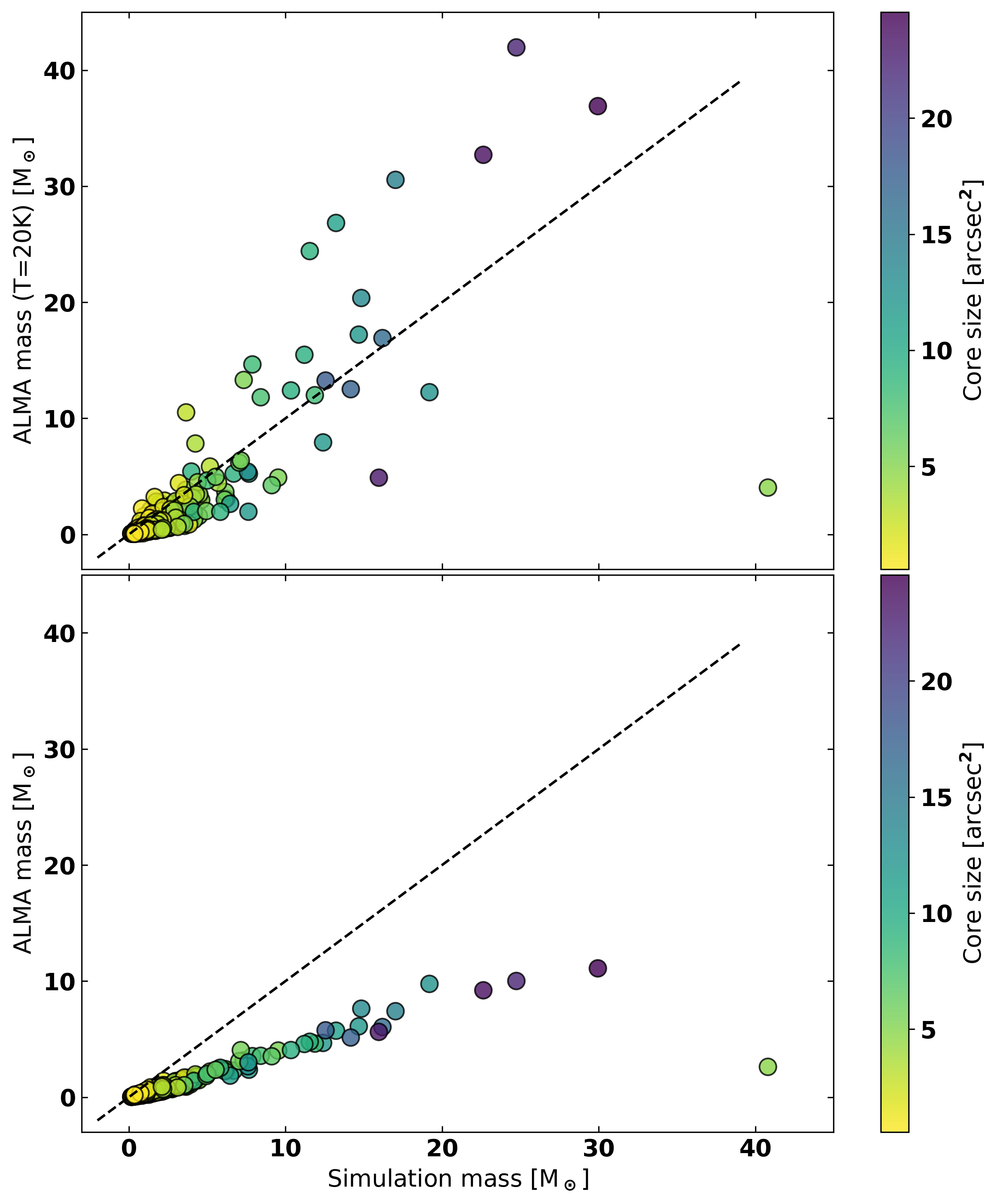}
    \caption{Comparison of the projected "simulation mass" of each mm core (the total mass within the core’s boundary derived
from the column density map of the original simulation before radiative transfer integrated along the line of sight, Section~\ref{sec:obs_v_sims}), and the core mass derived from the synthetic observations using (top:) the constant-temperature assumption T$_{\rm dust}$=20\,K and (bottom:) each core's intensity-weighted average temperature. The black dashed line shows a 1-1 correspondence. When using individual core temperatures, the mass derived from the synthetic observation is more tightly correlated with, but always lower than, the line-of-sight integrated simulation mass.}
    \label{fig:obs_v_sim_mass}
\end{figure}

Since we possess both the core masses derived from the synthetic observations, and the column densities directly from the simulation, it is possible to do a comparison of the fidelity with which they match. In Figure~\ref{fig:obs_v_sim_mass} we plot the total mass within each core's boundary derived from the \textsc{polaris} input grid before radiative transfer was applied compared to the core masses derived as described in Section~\ref{sec:core_evo}.

To derive the column mass of the cores from the original simulation, the input density grid is projected along the observer's viewing axis to make a column density map. The mass of each pixel is calculated as
\begin{equation}
    M = \mu_{H_2} m_H \int N_{H_2} dA \, ,
\end{equation}
\noindent where $\mu_{H_2}$ is the mean molecular weight per hydrogen molecule (equal to $2.8$), $m_H$ is the mass of a Hydrogen atom, and dA is the pixel area calculated as $dA = d^2 d\Omega$, where $d$ is the source distance and $d\Omega$ is the pixel solid angle.
The masses of the pixels within each core boundary are then summed to obtain the total projected "simulation mass" of each core.

Figure \ref{fig:obs_v_sim_mass} shows that, as might be expected, how the projected simulation masses compare to the masses derived from the synthetic observations depends on the temperature assumption adopted in calculating masses from the synthetic observations.  
For both of our temperature cases, there is a reasonably strong correlation between the mass derived from the synthetic observations and the projected simulation mass: the Pearson correlation coefficient is 0.81 (P-value=1.27e-117) for masses calculated using T$_{\rm dust}$=20\,K and 0.90 (P-value=1.44e-179) for masses calculated using the core intensity-weighted average temperatures.
While the correlation is stronger for core masses calculated using the intensity-weighted average core temperatures, it is notable that the pattern of the distribution of points differs in the two panels of Figure \ref{fig:obs_v_sim_mass}.
The masses derived from the synthetic observations assuming T$_{\rm dust}$=20\,K fall both above and below the 1:1 line, while the masses derived from the synthetic observations using intensity-weighted average core temperatures have less scatter and are always lower than the projected "simulation masses." 

Interestingly, the SED reference model analysis of \citet{Richardson24} similarly finds that core masses calculated from mm flux densities using a mass$\times$temperature-weighted average dust temperature (intended to approximate an intensity-weighted temperature) are well correlated with, but generally underestimate, the true masses calculated directly from their models (their Figure 7).  This is in contrast to core masses calculated from the mm flux densities using mass-weighted temperatures (which are not observationally accessible), which are even more tightly correlated but overestimate the true core masses \citep{Richardson24}.  As \citet{Richardson24} consider models with a fixed 1\,mm flux density within a given aperture, for a source at a given distance, the models shown in their Figure 7 have a maximum true mass of $\lesssim$2~M$_{\odot}$ and exhibit an inverse correlation between mass and temperature (i.e. the most massive cores have the lowest temperatures).  This is not the case in our work, where the more massive cores generally have higher temperatures (see Table~\ref{tab:properties} and Figure~\ref{fig:8panelhist}). 
As shown by the colourscale of the points in Figure \ref{fig:obs_v_sim_mass}, we also find that in our work the more massive cores -- which have masses of tens of M$_{\odot}$-- are generally larger in area.
We also note that, unlike the SED models of \citet{Richardson24}, our cores are embedded within a full molecular cloud.  There will thus be a substantial contribution from material along the line-of-sight, which likely contributes to the discrepancy between the projected simulation masses and the masses inferred from the synthetic observations.

\subsection{Transient cores in a globally collapsing high mass star-forming region}\label{sec:transient_cores}

As discussed in Section ~\ref{sec:core_numbers}, the mm continuum morphology of the region and the number of cores identified evolve significantly over the course of the simulation.
In the initial snapshots there is a single core that will eventually form the most massive sink in the simulation; as the region evolves, more cores become detectable. Notably, these cores are not distributed randomly but lie along `feeder-filaments' connecting to the central core. Both theoretical \citep[e.g.][]{Smith11a,Myers11,Smith16} and observational \citep[e.g][]{Kirk13,Peretto14,Motte18a} studies confirm that massive stars are likely to accrete along filaments. 

As the region evolves more cores are identified in the synthetic ALMA images due to i) the density of the whole region increasing, which contributes to making the mm continuum emission brighter, ii) new sinks forming and "lighting up" dust emission by increasing the dust temperature (see also Section~\ref{sec:core_numbers}), iii) fragmentation of the filament, and iv) global inflow along the filaments carrying new cores into the field of view. 
However, as discussed in Section~\ref{sec:core_numbers}, investigation of the boundaries of the cores in the feeder-filaments reveals that these boundaries are constantly shifting as the region evolves. This can be clearly seen in Figure \ref{fig:24paneldendro} where the blue lines representing the dendrogram core boundaries can shift significantly between snapshots even when the underlying emission appears broadly similar.
A good example of this is how the boundaries of the central core evolve with time. As cores travel along the feeder filaments towards the bottom of the potential well this alters the gas distribution at the centre of the region, which combines with observational effects such as the beamsize to cause different features to be identified as cores by the dendrogram analysis (see also Section~\ref{sec:core_numbers}).
The shifting core boundaries suggest that, unlike low mass cores sitting in isolation, many dendrogram-identifed cores in the synthetic ALMA images may represent clumpy peaks in the feeder-filaments rather than persistent structures.

Since the similarity of the functional forms of the core mass function and the stellar IMF has been known, it has often been theorised that the CMF \textit{determines} the IMF with some constant core efficiency \citep{Motte01,Alves07}. Using cores identified in 3D from simulation data using their gravitational potential, \citet{Smith09} showed that the correlation between the two distributions was statistical with a large scatter and did not apply to individual cores. As discussed in Section~\ref{sec:core2sink}, Figure~\ref{fig:CvSM} shows that core masses derived from our synthetic observations are also not related to the total mass of the enclosed sinks by a constant efficiency factor.
We can further investigate the correspondence between cores and `stars' using ~\ref{fig:maxSvC}, where the growth of the equivalent stellar mass of individual sinks is plotted against their millimetre core mass. 
As there is no mass-loss feedback tied to the sinks, they cannot become less massive with time, and so an increase in mass also means an increase in time (though not necessarily with a one-to-one mapping).
If core and star formation were separate sequential processes, the core mass would start high on the top left 
and then move steadily to the bottom right as the core mass drained into the sink. We do not see this trend in Figure~\ref{fig:maxSvC}, demonstrating that in the simulated massive star forming region core formation and star formation happen simultaneously, in a similar manner to the simulation-identified 3D cores in \cite{Smith09}. 

In contrast, an evolution from bottom left to top right on this plot would indicate that the sink was formed at almost the same time as the dendrogram-identified core but then accreted material more slowly than the core grew. There is evidence for this behaviour for the 5.9 M$_{\odot}$ sink associated with the core that we previously noted to be relatively isolated. However, for the most part there is little correspondence between the sink and core masses, showing the cores' fluctuating nature.

\begin{table}
    \setlength{\tabcolsep}{3pt}
	\centering
	\caption{Observables investigated as potential evolutionary indicators}
	\begin{threeparttable}
    \label{tab:obs_evol}
        \begin{tabular}{lcccc}
        \hline\hline
        Parameter & N$_{\rm core,min}^{a}$ & Snapshots$^{b}$ & Pearson & P-value \\
                 & & & Correl. Coeff. &\\
        \hline
        Peak intensity$^{c}$ & n/a & all & 0.72 &  7.05e-05\\[0.1cm]   
        Total flux$^{d}$ & n/a & all & 0.97 &  2.31e-15\\[0.1cm]    
        Number of cores & 1 & all & 0.96 &  7.20e-14\\[0.1cm]
        Total flux of cores$^{e}$ & 1 & all & 0.89 &  4.01e-09\\[0.1cm][0.1cm]
        $\frac{\rm Flux~in~cores}{\rm Total~flux}$ & 1 & all & -0.59 & 0.0025 \\[0.1cm]
        ($\Delta$l$_{\rm mean}$)$^{f}$ & 2 & $\ge$171& 0.64 &  0.001 \\[0.1cm]
        (l$_{\rm min})^{g}$ & 15 & $\ge$178 & -0.31 &  0.26 \\[0.1cm]
        Q & 15 & $\ge$178 & 0.54 &  0.04 \\[0.1cm]
        \hline\hline
    \end{tabular}
    \begin{tablenotes}
    \item[$a$] Minimum number of cores required to calculate the parameter.
    \item[$b$] Snapshot numbers for which the parameter can be calculated, see Table~\ref{tab:snapshot_stats}.
    \item[$c$] Peak intensity within the field of view, measured from images corrected for the primary beam response. This is the same as the peak intensity of the brightest mm core (Section~\ref{sec:core_numbers}).
     \item[$d$] Total flux density of $\ge$5$\sigma$ emission within the 30\% mosaic response, measured from images uncorrected for the primary beam response, using the CASA {\sc imstat} task.  
     \item[$e$] Total flux density in cores = the sum of the integrated flux densities of the cores identified with {\sc astrodendro}, measured using {\sc astrodendro} from images uncorrected for the primary beam response.           
     \item[$f$] Mean projected separation of cores identified with {\sc astrodendro} from the brightest mm core.
      \item[$g$] Minimum projected distance between cores identified with {\sc astrodendro}, measured from the minimum spanning tree (MST).
    	\end{tablenotes}
	\end{threeparttable}
\end{table}

\subsection{Observational indicators of evolutionary state?}\label{sec:obs_evol}

The goal of this work is to gain insight into how the observable mm continuum properties of massive star forming regions evolve in time, something which is obviously impossible in a single observation.
Using our synthetic ALMA images, we investigated the time evolution of a range of readily measurable properties, with the aim of identifying any observables that could be used as indicators of  \emph{relative} evolutionary stage for observed regions.

The parameters we investigate, listed in Table~\ref{tab:obs_evol},  focus on properties that are commonly measured from mm interferometric observations and/or that have been suggested in the literature as potential indicators of evolutionary state.  In addition to the number of cores and the total 1.3\,mm flux density of $\ge$5$\sigma$ emission within the field of view \citep[similar to the S$^{\rm recovered}_{\rm 1.3mm}$ parameter from][]{almaimf1},  discussed in Section ~\ref{sec:core_numbers}, we consider the peak intensity within the field of view (which is the same as the peak intensity of the brightest mm core, Section ~\ref{sec:core_numbers}), the total 1.3\,mm flux density of identified cores, and the ratio of the total flux density in cores to the total flux density within the field of view.  The latter property is comparable to the $M^{\rm cores}_{\rm 1.3 mm}/M^{\rm recovered}_{\rm 1.3mm}$ mass ratio from \citet{almaimf1}, given their assumption of a constant T$_{\rm dust}$=20\,K for this calculation.  We also investigate three parameters related to protocluster structure: the mean separation of other cores from the brightest mm core \citep[e.g.][]{Cyganowski17}, the minimum separation between cores \citep[e.g.][]{Traficante23}, and the $\mathcal{Q}$ parameter \citep{Cartwright2004}.  The $\mathcal{Q}$ parameter, in particular, has been found to increase (corresponding to increasing central condensation) with time in simulations of hierarchical star cluster formation when calculated using sinks \citep{Maschberger10} and to correlate with clump-scale L/M when calculated using cores \citep{Xu24}, suggesting it as a possible evolutionary indicator.

The minimum separation between cores (l$_{\rm min}$ in Table~\ref{tab:obs_evol}) and the $\mathcal{Q}$ parameter  are calculated using
Minimum Spanning Trees \citep[MSTs;][]{Barrow85} -- defined as the set of straight edges which connect all the points in a plane with the minimum total edge lengths and without forming any closed loops -- which we compute from the core distribution in each snapshot with the Python package MiSTree \citep{Naidoo2019}. \citet{Cartwright2004} derive the $\mathcal{Q}$ parameter from the MST, which quantifies the structure of the distribution:

\begin{equation}
\mathcal{Q} = \frac{\bar{m}}{\bar{s}} = \frac{m/\biggr(\sqrt{\frac{N_\text{tot} \pi R_{\text{clus}}^2}{(N_\text{tot}-1)}}\biggr)}{s/R_\text{clus}},
	\label{eq:q}
\end{equation}

\noindent where $\bar{m}$ is the mean edge length, normalised by the total number of points in the distribution $N_\text{tot}$ and the cluster radius $R_\text{clus}$. $\bar{s}$ is the correlation length, which is defined as the mean pairwise separation between points, normalised by $R_\text{clus}$. $R_\text{clus}$ is defined as the distance from the mean position of all points to the point furthest from this mean position. 
Due to the caveats associated with calculating $\mathcal{Q}$ for small source counts \citep[e.g. Appendix B of][]{Avison2023}, we compute $\mathcal{Q}$ and l$_{\rm min}$ only for snapshots with at least 15 cores, as indicated in Table~\ref{tab:obs_evol}.  

The time evolution of all investigated parameters is shown in Figure~\ref{fig:parameters_v_time}; Pearson correlation coefficients\footnote{We use scipy.stats.pearsonr to compute the Pearson correlation coefficient, which measures the degree of linear correlation between two variables. The command also returns the p-value, which is the probability the same correlation could be obtained from randomly distributed data. A value of p$<$0.05 is conventionally taken to indicate a statistically significant correlation, i.e. the rejection of the null hypothesis that there is no correlation between the two variables being considered.}, computed over all timesteps that satisfy our criterion for the minimum number of cores, are presented in Table~\ref{tab:obs_evol}.  It is clear from Table~\ref{tab:obs_evol} that the parameters that most strongly correlate with time are the number of cores (see Section~\ref{sec:core_numbers}) and the total mm flux density within the field of view, followed by the total flux of the identified cores.  The peak intensity exhibits a weaker, but still strong, overall correlation with time.  However, as discussed in Section ~\ref{sec:core_numbers}, there are large fluctuations timestep-to-timestep and periods of the simulation when the correlation is poor. Similarly, while the ratio of the total flux density in cores to the total flux density in the field of view shows a statistically significant moderate negative correlation with time, this ratio fluctuates notably from timestep to timestep (see Figure~\ref{fig:parameters_v_time}). This is largely due to the rapidly shifting core boundaries discussed in Section~\ref{sec:transient_cores}. Interestingly, \citet{almaimf1} remark on the large relative dispersion of $M^{\rm cores}_{\rm 1.3 mm}/M^{\rm recovered}_{\rm 1.3mm}$ in the ALMA-IMF sample (22\% $\pm$13\%, c.f. Figure~\ref{fig:parameters_v_time}), but find no significant differences between subsamples grouped by evolutionary stage. 

The $\mathcal{Q}$ parameter shows only a moderate Pearson correlation coefficient and just meets the significance threshold. There is a mild increase in this parameter throughout the evolution, but $\mathcal{Q}$ remains below the 0.8 value which is expected for dynamically unevolved objects with substructure \citep{Cartwright2004}. This is in contrast to the work of \citet{Maschberger10} who find that in molecular cloud protostellar sub-clusters, $\mathcal{Q}$ evolves to higher values representative of more centrally condensed distributions. This difference is likely attributable to the region imaged here being only the central part of a larger clump. New material flows in along the filament from outside the imaged region meaning $\mathcal{Q}$ stays low as appropriate for an extended region with sub-structure.

l$_{\rm min}$ shows a negative correlation with time. Whilst the correlation does not have a p-value small enough to be considered statistically significant, \citet{Traficante23} similarly observe a negative correlation between l$_{\rm min}$ and the evolutionary state of their clusters (as traced by clump $L/M$). Such a result is in support of a dynamical, clump-fed massive star formation scenario in which the global infall of material results in the formation of more cores with closer separations at later times. Notably, we do not see any indication of the computed l$_{\rm min}$ being limited by the angular resolution of the synthetic observations.  In angular units, the smallest measured l$_{\rm min}$ is 1\farcs12, $\sim$1.6$\times$ the geometric mean of the synthesised beam (0\farcs81$\times$0\farcs62), while the largest measured l$_{\rm min}$ is 2\farcs18.
An additional signature of global infall in this simulation is illustrated by the statistically significant positive correlation of the mean distance of cores from the brightest mm core with time. As material is channelled into the clump from larger scales, the number of cores increases, the cluster is built up across a larger portion of the imaged region, and the mean distance from the most massive core increases.

In summary, while we tested a variety of observational indicators, none of the parameters are particularly strongly correlated with time other than total 1.3\,mm flux density, total 1.3\,mm flux density in cores, and the number of cores.
These indicators show positive correlations but are also likely to be affected by differences between regions other than age (such as total mass). Future work would be needed using a range of different initial conditions to see if these measurements would still be indicative of evolutionary stage when applied to a diverse sample of regions as might be seen in an observational survey.

\subsection{Caveats and future work}\label{sec:caveats}
The analysis presented here most likely represents an upper limit on the amount of fragmentation in such regions as i) the feedback is post-processed and hence does not dynamically affect the gas, and ii) the simulations lack magnetic fields. These effects are likely to reduce the number of fragments formed \citep[e.g.][]{Rosen19}. However, we do see that the core boundaries fluctuate from the very beginning of the simulation before feedback is significant. Future work should repeat this study but with feedback included in the underlying simulation.

In order to model the feedback, assumptions had to be made about the protostellar sources. We used the MIST database to parameterise this effect with a fixed sink-to-star efficiency of 50\%. This approach provides a reasonable estimate based on the work of \citet{Konyves15}, but naturally if the efficiency were higher the sources would be more luminous and vice-versa. Similarly, assumptions had to be made about the dust model, however, we took care to ensure that our $\kappa$ value was around 1\,cm$^2$\,g$^{-1}$ as would be expected (see Section~\ref{sec:RT}), and confirmed that this leads to intensity values in the expected range. 

Source identification in interferometric mm images is itself an active area of research, with at least four major algorithms currently in use in the literature (\textit{Astrodendro}, \citealt{Rosolowsky08dendro}; \textit{CuTEx}, \citealt{Molinari2011}; \textit{Hyper}, \citealt{Traficante2015}; and \textit{getsf}, \citealt{Menshchikov2021}; see also discussion in \citealt{Coletta_almagal3}.)  We choose to use \textsc{astrodendro} in our analysis because it has been, and continues to be, commonly employed to identify structures in ALMA images of young massive star-forming regions, including samples of hub-filament systems \citep[e.g.][]{Anderson2021}, ALMA-IRDC \citep[e.g.][]{Barnes2021}, and the ALMA Survey of 70$\mu$m Dark High-mass Clumps in Early Stages \citep[ASHES;][]{Sanhueza19,Morii23,Morii2024}.  The longevity of \textsc{astrodendro} as a publicly-available state-of-the-art tool also means that it is commonly used as a reference-algorithm comparison for other source extraction approaches \citep[e.g.][]{Traficante23,Coletta_almagal3,Cheng24}, and these comparisons have demonstrated that different algorithmic approaches identify different core populations.  For example, \citet{Cheng24} find that in the Investigations of Massive Filaments and Star Formation (INFANT) sample, \textsc{astrodendro} identifies significantly more cores than \textit{getsf}, with both algorithms identifying bright, high-contrast cores but \textsc{astrodendro} also identifying a population of weak "cores" with irregular boundaries.
This is consistent with our finding that many of the "cores" identified by \textsc{astrodendro} in our synthetic observations are transient features in the filamentary accretion flows.  Future work repeating our analysis of the synthetic observations using different source extraction algorithms would be needed to establish the level of transience of cores identified using non-hierarchical approaches.

A further caveat is that we model a single region in this analysis and use it as a test case to examine the temporal evolution. We therefore cannot say that every massive star formation region evolves in this manner, but we can say that it should not be \textit{automatically} assumed that cores in these regions are well defined objects that persist over the massive YSO's lifetime. It should also be emphasised that these results apply to collapsing high mass regions. In more isolated low mass regions where there is less interaction there may be a closer correspondence between core mass and star mass.

\section{Conclusions}\label{sec:conclusions}

In this work, we have leveraged the ability of synthetic observations to probe the time evolution of high-mass star-forming regions to investigate how the properties and protostellar content of cores that would be identified by a dendrogram analysis of 1.3\,mm ALMA images evolve with time, and to test proposed mm-wavelength observational indicators of relative evolutionary state. To do this, we generated a time series of synthetic ALMA 12m array images by post-processing an AREPO simulation using the \textsc{Polaris} code to perform dust radiative transfer and then using CASA to create and image synthetic visibility data. 
Our main findings are as follows.
\begin{itemize}
    \item The total 1.3\,mm flux density of the region and the number of cores identified by the dendrogram analysis increase with time. Initially, only a single core is seen, associated with what will become the most massive star, but as more gas flows into the region, and this flow becomes denser and more fragmented, more structures are identified. 

    \item The 1.3\,mm peak and integrated intensity of the brightest mm core in our synthetic ALMA images do not increase monotonically as the most massive sink in this core grows in mass.  This indicates it cannot be assumed that brighter mm sources host more massive protostars.

   \item The majority of the dendrogram-identified cores are transient features whose boundaries change from snapshot to snapshot, as the massive star at the centre of the region forms. Rather than being distinct objects, these cores may represent clumpy peaks in a feeder-filament system.

    \item Using the output of the \textsc{Polaris} radiative transfer to calculate an observationally-relevant intensity-weighted average temperature for each core, we find that even at relatively early times (t=3.896 Myr) the distribution of core temperatures spans a range of tens of Kelvin.  The first core with an intensity-weighted average temperature $>$20\,K appears when the effective mass of the most massive sink in the simulation is only 1.695M$_{\odot}$ (at t=3.765 Myr) and the width of the core temperature distribution increases with time.

    \item Comparing the evolution of core mass functions (CMFs) calculated using individual core temperatures and using the common observational assumption of a single temperature for all cores within a clump (for which we adopt a fiducial value T$_{\rm dust}$=20~K), we show that this single-temperature assumption overestimates the masses of high-mass cores and underestimates the masses of some low-mass cores.  This artificially broadens the core mass distribution.

    \item Most cores identified by our dendrogram analysis of the synthetic ALMA 1.3\,mm images do not contain any sinks within their boundaries.  Those that do generally contain a small number $n \leq3$, with the exception of the central, brightest mm core, which typically contains $n>10$ once the region becomes evolved. 
    
    \item The total sink mass within each dendrogram-identified core is not strongly correlated with the mass of the core calculated using its intensity-weighted average temperature. For low total sink mass the core mass is normally larger than a 1-1 correspondence, while for higher total sink masses ($\gtrsim$2M$_{\odot}$) the core mass is typically lower than the total sink mass.  For individual sinks it is difficult to connect core mass to the final sink mass as, apart from a single isolated core, the masses of the host cores v. time resemble a random walk as the sinks grow in mass.

 \item The relationship between "true" core masses derived from the simulation column density before radiative transfer and the core masses derived from the synthetic ALMA observations depends strongly on the temperature assumption used in deriving masses from the synthetic dust emission.  Simulation and synthetic-observation derived masses are correlated for both temperature assumptions (individual intensity-weighted average core temperatures and a constant T$_{\rm dust}$=20\,K). The relationship for core masses derived using the individual core temperatures shows a tighter (less scatter) and stronger correlation (Pearson correlation coefficient 0.90).  The masses derived from the synthetic ALMA observations using individual core temperatures are however increasingly lower than the simulation masses as core mass and size increase.  In contrast, when core masses are derived assuming T$_{\rm dust}$=20~K, the relation between the simulation and synthetic observation-derived masses has a much larger scatter, with masses derived from the synthetic observations in many cases larger than the simulation mass.

    \item We measured eight observational properties as potential indicators of evolutionary state, including cluster structure metrics derived from minimum spanning trees (MSTs), and tested these properties for correlation with time.  The only properties to show a strong (Pearson correlation coefficient $\geq$0.89), statistically significant correlation with time are the total 1.3\,mm flux density of the region, the total 1.3\,mm flux density in cores, and the number of cores.  As these indicators are also likely to be affected by differences other than age (such as the total mass of a region or the imaged field of view), our results indicate  that 1.3\,mm properties alone cannot be readily used to track relative evolutionary state.

\end{itemize}

\section*{Acknowledgements}
We thank an anonymous referee for helpful comments that improved this paper.  We gratefully acknowledge Glen Hunter whose work during his MPhys thesis was the starting point for much of this analysis.
RJS gratefully acknowledges an STFC Ernest Rutherford fellowship (grant ST/N00485X/1) and CJC gratefully acknowledges support from the STFC (grant ST/Y002229/1).
This work used the DiRAC@Durham facility managed by the Institute for Computational Cosmology on behalf of the STFC DiRAC HPC Facility (www.dirac.ac.uk). The equipment was funded by BEIS capital funding via STFC capital grants ST/P002293/1, ST/R002371/1 and ST/S002502/1, Durham University and STFC operations grant ST/R000832/1. DiRAC is part of the National e-Infrastructure.  The National Radio Astronomy Observatory is a facility of the National Science Foundation operated under cooperative agreement by Associated Universities, Inc.
This research made use of astrodendro, a Python package to compute dendrograms of Astronomical data (\url{http://www.dendrograms.org/}), 
Astropy: \url{http://www.astropy.org}, a community-developed core Python package and an ecosystem of tools and resources for astronomy \citep{astropy:2013, astropy:2018, astropy:2022}, and APLpy, an open-source plotting package for Python \citep{aplpy2012, aplpy2019}.

\section*{Data Availability}
The unprocessed radiative transfer simulations prior to the creation of the synthetic ALMA observations or core identification are available on the Zenodo website (DOI 10.5281/zenodo.17351347). Online tables provide the full catalogue of synthetic core properties.



\bibliographystyle{mnras}
\bibliography{bib}



\onecolumn

\appendix

\section{\textsc{polaris} 1.3 mm images}
\label{App1}
\begin{figure*}
    \centering
    \includegraphics[width=0.97\linewidth]{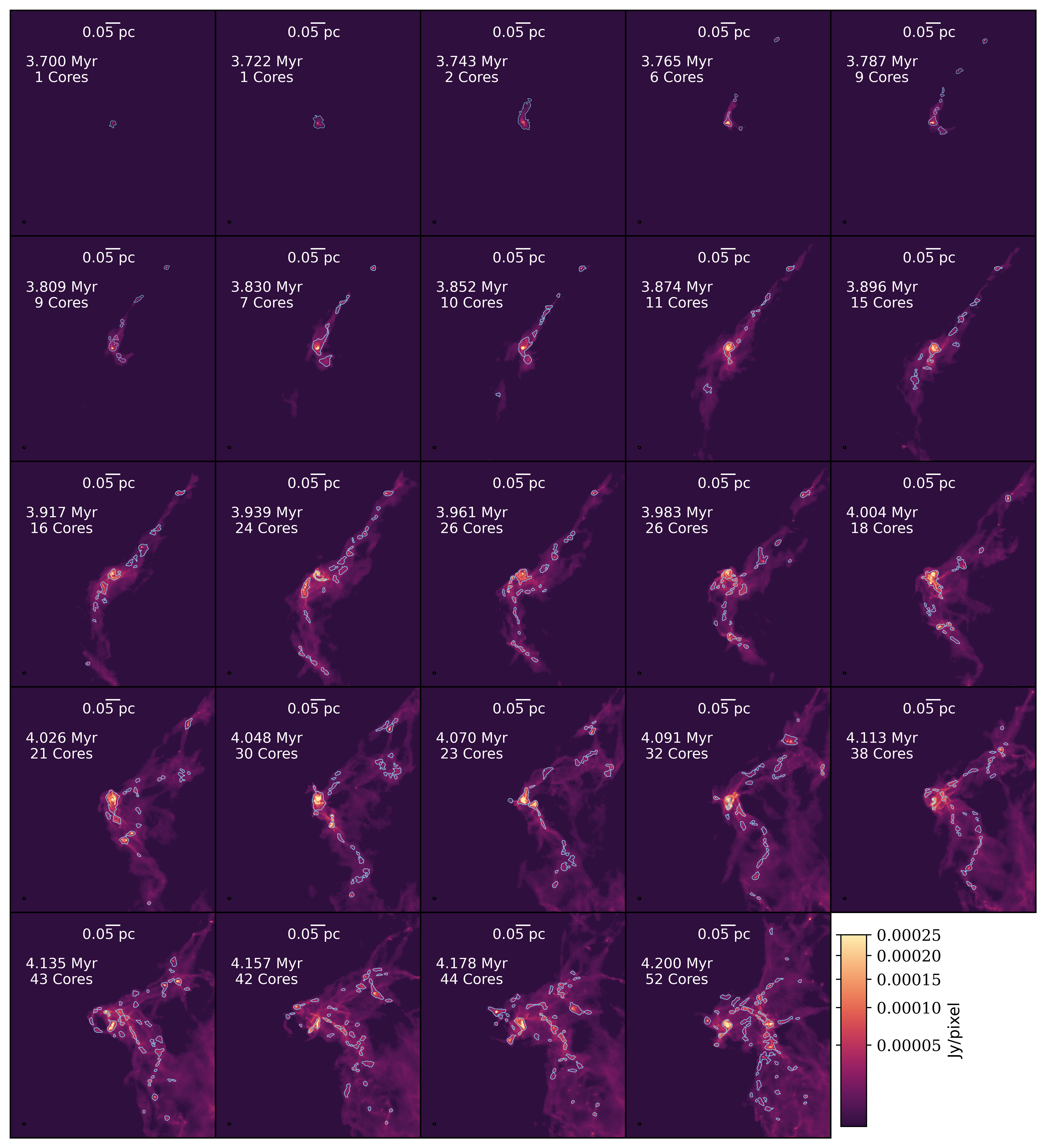}
    \caption{\textsc{polaris} 1.3\,mm images for each snapshot (Section~\ref{sec:RT}), centred on the position of the most massive sink and with the core boundaries identified with \textsc{astrodendro} from the corresponding synthetic ALMA 1.3\,mm images (Section~\ref{sec:core_ident}, Table~\ref{tab:properties}) overlaid in blue. The colourscale and field of view are the same for all images, and the scalebars assume D=2\,kpc (Section~\ref{sec:RT}). In each panel, the time of the snapshot in Myr and number of identified cores are given at upper left.}
    \label{fig:polaris_images}
\end{figure*}

Figure~\ref{fig:polaris_images} shows the \textsc{polaris} 1.3\,mm images for each of the simulation snapshots.

\section{Synthetic ALMA 1.3 mm images}
\begin{figure*}
    \centering
    \includegraphics[width=0.97\linewidth]{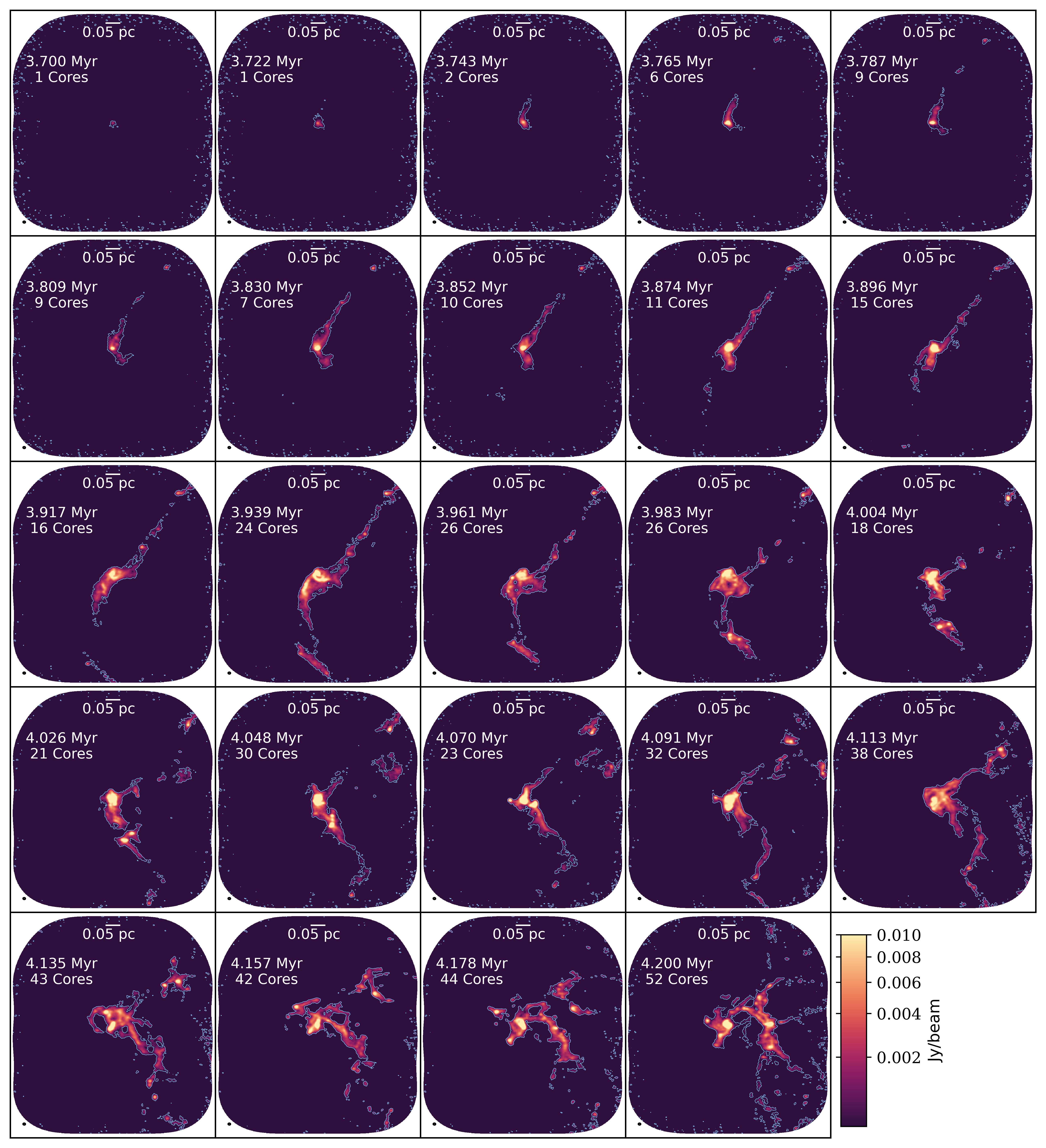}
    \caption{Same as Figure~\ref{fig:24paneldendro} but overlaid with 5$\sigma$ contours of the synthetic ALMA 1.3\,mm continuum images, to highlight the extent of detectable emission.  The images shown have been corrected for the primary beam response. In each panel, the synthesised beam is shown at lower left and the time of the snapshot in Myr and number of identified cores are given at upper left.}
    \label{fig:24panelsigma}
\end{figure*}

Figure~\ref{fig:24panelsigma} shows the synthetic ALMA 1.3\,mm continuum images as in Figure ~\ref{fig:24paneldendro}, but overlaid with contours of the synthetic continuum emission rather than the boundaries of the {\sc Astrodendro} cores.

\section{Observables investigated as potential evolutionary indicators}

\begin{figure*}
    \centering
    \includegraphics[width=0.35\textwidth]{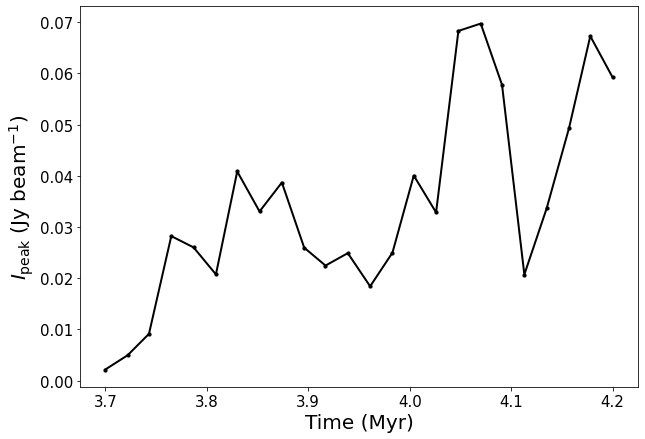}\includegraphics[width=0.35\textwidth]{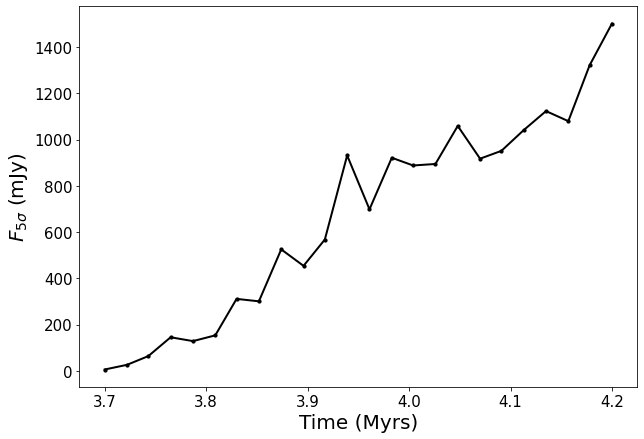}
    \includegraphics[width=0.35\textwidth]{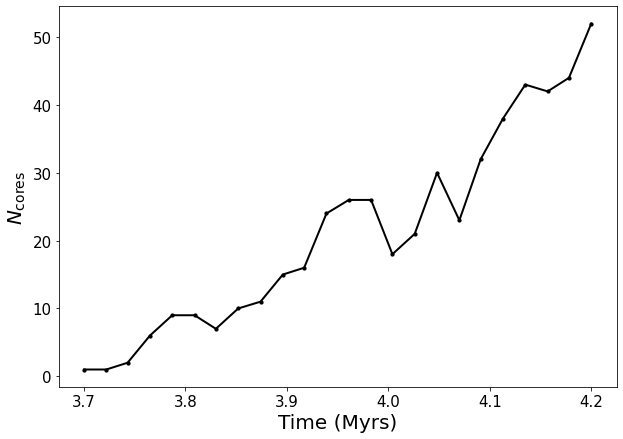}\includegraphics[width=0.35\textwidth]{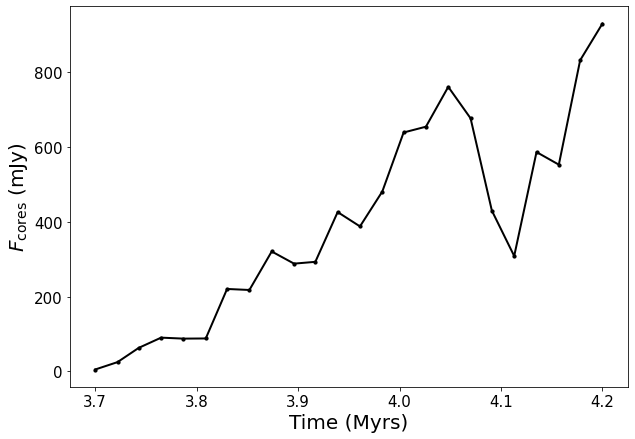}
    \includegraphics[width=0.35\textwidth]{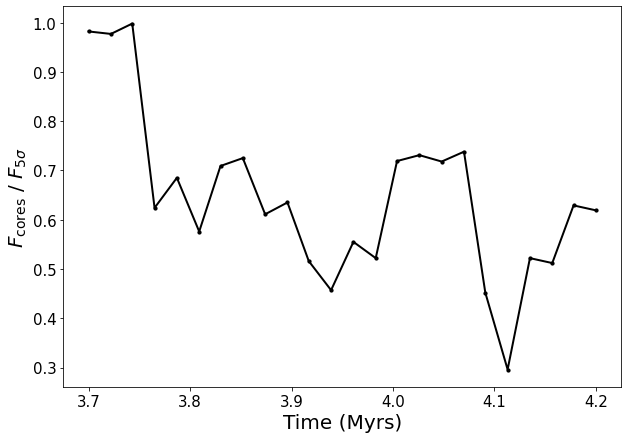}\includegraphics[width=0.35\textwidth]{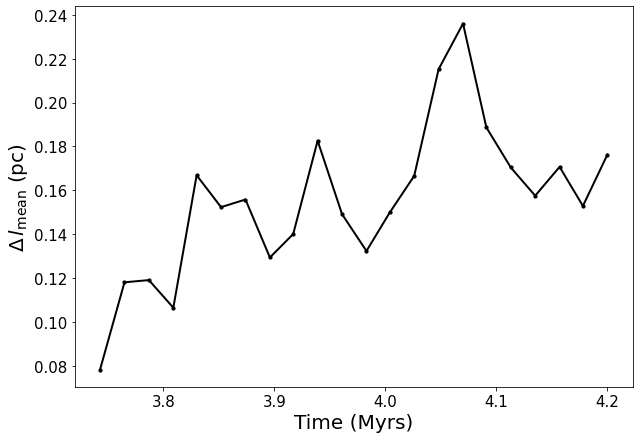}
    \includegraphics[width=0.35\textwidth]{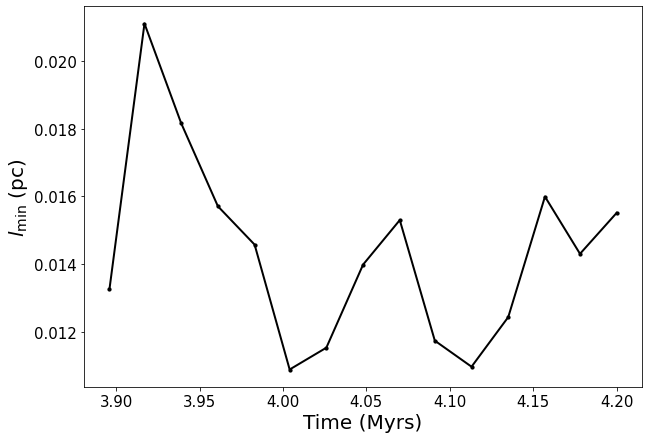}\includegraphics[width=0.35\textwidth]{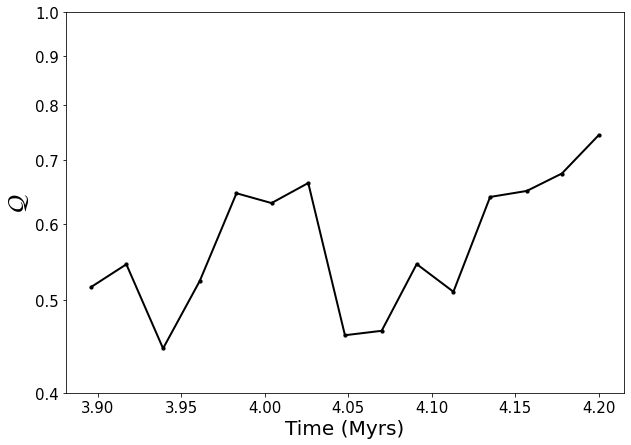}
    \caption{Time evolution of parameters in Table~\ref{tab:obs_evol}.  Top row: Left: Peak intensity ($I_\text{peak}$), measured from images corrected for the primary beam response.  Right: Total flux density of $\ge$5$\sigma$ emission within the 30\% mosaic response ($F_{5\sigma}$), measured from images uncorrected for the primary beam.  Second row: Left: Number of cores identified with {\sc astrodendro} ($N_\text{cores}$).  Right: Sum of core flux densities, measured from images uncorrected for the primary beam response ($F_\text{cores}$).  Third row: Left: Ratio of flux in cores to total flux density of $\ge$5$\sigma$ emission within the 30\% mosaic response ($F_\text{cores}/F_{5\sigma}$).  Right: Mean projected separation of cores identified with {\sc astrodendro} from the brightest mm core ($\Delta \, l_\text{mean}$) .  Bottom row: Left: Minimum projected separation between cores ($l_\text{min}$). The size of the synthesised beam is 0.0078$\times$0.0060 pc at D=2kpc.  Right: $\mathcal{Q}$.  Note that x-axis ranges differ between panels based on the number of snapshots for which each parameter can be calculated, see Table~\ref{tab:obs_evol}.   }
    \label{fig:parameters_v_time}
\end{figure*}

Figure~\ref{fig:parameters_v_time} shows the time evolution of observable properties investigated as potential indicators of evolutionary stage.  Pearson correlation coefficients and p-values are given in Table~\ref{tab:obs_evol}.


\bsp	
\label{lastpage}
\end{document}